\documentclass[5p]{elsearticle}
\usepackage{amssymb}
\usepackage{amsmath}
\usepackage{multicol}
\usepackage{graphicx}
\usepackage{multirow}
\usepackage{textcomp}
\usepackage{rotating}
\usepackage{array}
\usepackage{subfigure}
\newtheorem{definition}{Meta Definition}
\newtheorem{procedure}{Meta Procedure}

\begin{document}

\title{A Classification for Community Discovery Methods in Complex Networks}

\author[csd,kdd,bal]{Michele Coscia}
\ead{coscia@di.unipi.it}

\author[kdd]{Fosca Giannotti}
\ead{fosca.giannotti@isti.cnr.it}

\author[csd,kdd]{Dino Pedreschi}
\ead{pedre@di.unipi.it}

\address[csd]{Computer Science Department, University of Pisa, Pisa, Italy}
\address[kdd]{KDDLab, ISTI-CNR, Pisa, Italy}
\address[bal]{Barabasi Lab, Northeastern University, Boston, USA}

\begin{abstract}
In the last few years many real-world networks have been found to show a so-called community structure organization. Much effort has been devoted in the literature to develop methods and algorithms that can efficiently highlight this hidden structure of the network, traditionally by partitioning the graph. Since network representation can be very complex and can contain different variants in the traditional graph model, each algorithm in the literature focuses on some of these properties and establishes, explicitly or implicitly, its own definition of community. According to this definition it then extracts the communities that are able to reflect only some of the features of real communities. The aim of this survey is to provide a manual for the community discovery problem. Given a meta definition of what a community in a social network is, our aim is to organize the main categories of community discovery based on their own definition of community. Given a desired definition of community and the features of a problem (size of network, direction of edges, multidimensionality, and so on) this review paper is designed to provide a set of approaches that researchers could focus on. 
\end{abstract}


\maketitle

\section{Introduction}

A complex network is a mathematical model of interaction phenomena that take place in the real world, which has revealed a powerful computational basis for the analysis of such phenomena.
One critical problem, which has been widely studied in the literature since the early analysis of complex networks, is the identification of communities hidden within the structure of these networks. 

A {\em community} is intuitively understood as a set of entities where each entity is closer, in the network sense, to the other entities within the community than to the entities outside it. Therefore, communities are groups of entities that presumably share some common properties and/or play similar roles within the interacting phenomenon that is being represented. Community detection is important for many reasons, including node classification which entails homogeneous groups, group leaders or crucial group connectors. Communities may correspond to groups of pages of the World Wide Web dealing with related topics \cite{flake-flow}, to functional modules such as cycles and pathways in metabolic networks \cite{metabolic-community, kclique}, to groups of related individuals in social networks \cite{edgebetween} and so on.

Community discovery has analogies to the clustering problem, a traditional data mining task. In data mining, clustering is an unsupervised learning task, which aims to partition large sets of data into homogeneous groups (clusters). In fact, community discovery can be viewed as a data mining analysis on graphs: an unsupervised classification of its nodes. In addition, community discovery is the most studied data mining application on social networks. Other applications, such as graph mining \cite{gspan}, are still in an early phase of their development. Instead community discovery has achieved a more advanced development with contributions from different fields, such as statistical physics.

Nevertheless, this is only part of the community discovery problem. In classical data mining clustering, we have data that is not in a relational form. Thus, in this general form, the fact that the entities are nodes connected to each other through edges has not been thoroughly explored. Spatial proximity needs to be mapped to network proximity between entities represented as vertices in a graph.

The most accepted definition of proximity in a network is based on the topology of its edges. In this case the definition of community is formulated according to the differences in the densities of links in different parts of the network. Many networks have been found to be non-homogeneous, consisting not of an undifferentiated mass of vertices, but of distinct groups. Within these groups there are many edges between vertices, but between groups there are fewer edges. The aim of a community detection algorithm is, in this case, to divide the vertices of a network into some number $k$ of groups, while maximizing the number of edges inside these groups and minimizing the number of edges established between vertices in different groups. These groups are the desired communities of the network.

This definition reveals vague and unprecise as the complexity of network representations increases and novel analytical settings emerge, such as information propagation or multidimensional network analysis. For example, in a temporal evolving setting, two entities can be considered close to each other if they share a common action profile even if they are not directly connected. Often times, a novel approach to community discovery is designed to face a specific problem and it has developed its own definition of community.

In addition to the variety of different definitions of community, communities have a number of interesting features. They can exhibit a hierarchical or overlapping configuration of the groups inside the network. Or else the graph can include directed edges, thus giving importance to this direction when considering the relations between entities. The communities can be dynamic, i.e., evolving over time, or multi-relational, i.e., there could be multiple relations and sets of individuals that behave as isolated entities in each relation of the network, thus forming a dense community when considering all the possible relations at the same time.

As a result this extreme richness of definitions and features has lead to the publication of an impressive number of excellent solutions to the community discovery problem. It is therefore not surprising that there are a number of review papers describing all these methods, such as \cite{bfsurvey}. 

We believe that a new point of view is needed to organize the body of knowledge about community detection, shifting the focus from {\em how} communities are detected to {\em what} kind of communities are we interested to detect. Existing reviews tend to analyze the different techniques from a procedural perspective. They cluster the different algorithms according to their operational method, not according to the definition of community they adopt in the first place. Nevertheless, there are many different ways to conceive a community within a network, as acknowledged also by \cite{leicht-bayes-def}, where authors maintain that ``[all the methods] require us to know what we are looking for in advance before we can decide what to measure'' --- here ``know what we are looking for'' clearly means to define what a community really is. To use a metaphor, existing reviews talk about the bricks and mortar that make up a building with no mention about its architectural style. In other words, the aim of the previous reviews is to talk to people interested in building a new community detection algorithm, rather than those who want to use the methods presented in the literature. Our aim is precisely the latter.

We have thus chosen to cluster the community discovery algorithms by considering their reference definition of what is a community, which depends on what kinds of groups they aim to extract from the network. For each algorithm we record the characteristics of the output of the method, thus highlighting which sets of features the reviewed algorithm is suitable or not suitable for. We also consider some general frameworks that provide both a community discovery approach and a general technique. These are applicable to other graph partitioning algorithms by adding new features to these other methods.

The paper is organized as follows. In Section 2 we provide a general definition of the community discovery problem and the meta definition of what a community is. In Section 3 we explain the classification of algorithms based on community definitions. Then, from Sections 4 to 12, we present the main categories of approaches given our problem definition, along with what we consider to be the most important works in each given category. In Section 13 we provide various evaluation measurements over a collection of reviewed methods on a benchmark graph. Section 14 reviews some other related works, reviews regarding community discovery in social networks, along with the rationale behind the novel approach to these methods provided in this paper. Finally, Section 15 concludes the survey and provides an approach to possible future work.

\section{Problem Definition}\label{sec:probdef}

\subsection{Problem Representation}\label{sec:probrep}
Let us assume that we have a graph $G$ denoted by a quadruple $G=(V,E,L,C)$, where $V$ is a set of labeled nodes, $E$ is a set of labeled edges, $L$ is a set of edge labels and $C$ is a set of node labels. $E$ is a set of quadruples of the form $(u,v,l,w)$ where $u,v \in V$ are nodes, $l \in L$ is a label and $w$ is an integer that represents the weight of the relation. We assume that given a pair of nodes $u,v \in V$ and a label $l \in L$ only one edge $(u,v,l,w)$ may exist; however the direction of the edge is considered in the model, thus edges $(u,v,l,w)$ and $(v,u,l,w)$ are considered distinct. We also assume that each node can be labeled with one or more category $c \in C$. In addition, we consider the temporal evolution of the network. Thus each edge, and node, can be labeled with an arbitrary number of timestamps that represent the time in which the edge appears and disappears in the network. The labels of a given node can also change over time. Note that nodes can create/delete edges in the network and/or change/introduce/delete one or more labels in their category set. We call such events ``actions'' that are performed by the nodes.
 
With this complex model we can represent all possible variants in a graph of a complex real world phenomenon. For example, we can model multi-relational networks by considering the edge labels $L$ as the different relations (dimensions) of the network. We can also represent simpler models, such as unweighted networks, by assigning the same weight $w=1$ to every edge in the network.
Hereafter we will use the notation presented in Table 1. We introduce new symbols and notations when they are needed for the presentation of one particular method but not useful for the others.

\begin{table}
\begin{center}
\begin{tabular}{|c|c|}
\hline
Symbol & Description \\
\hline
n & Number of vertices of the network\\
m & Number of edges of the network\\
k & Number of communities of the network\\
$\bar{K}$ & Avg degree of the network\\
K & Max degree in the network\\
T & Number of action in the network\\
A & Max number of actions for a node\\
D & Number of dimensions (if any)\\
c & Number of vertex types (if any)\\
t & Number of time step (if any)\\
\hline
\end{tabular}
\end{center}
\label{tab:notation}
\caption{Resume of the main notation used in the paper.}
\end{table}

\subsection{Community Meta Definition}\label{sec:metadef}
We will now present our meta definition of a community in a complex network. With this meta definition we create an underlying concept which is the basis behind this survey and includes all the possible definition variants present in the literature.

\begin{definition}[Community]\label{def:community}
A community in a complex network is a set of entities that share some closely correlated sets of actions with the other entities of the community. Here we consider direct connection as a particular, and very important, kind of action. 
\end{definition}

The aim of a community discovery algorithm is to identify these communities in the network. The desired result is a list of sets of entities grouped together. Starting from this meta definition we can model the main aspects of  discovering communities in complex networks.

\textbf{Density-based definitions}. In this classical setting, as we mentioned in the Introduction, the definition is entirely based on the topology of the network edges. The community is defined as a group in which there are many edges between vertices, but between groups there are fewer edges. The aim of a community detection algorithm is to divide the vertices of a network into some number $k$ of groups, while maximizing the number of edges inside these groups and minimizing the number of edges that run between vertices in different groups. In our definition we consider the connection between two vertices a particular kind of action. Hence, if we group entities by maximizing their common actions, we also group them by maximizing the edges inside the community. Community discovery is exactly the same if the edge creation is the only action recorded in the network representation. In addition, by considering different kinds of sets of action in the meta definition, we can also model the overlapping situation: for certain sets of actions (i.e. connections) a node belongs to one community, for another set of actions, it belongs to another community.

\textbf{Vertex similarity-based definitions}. As pointed out by Fortunato \cite{bfsurvey}, it is natural to assume that communities are groups of vertices that are similar to each other. One can compute the similarity between each pair of vertices with respect to some reference property, local or global, irrespectively of whether or not they are connected by an edge. Each vertex ends up in the cluster whose vertices are the most similar to it. By considering an evolving setting in our problem representation, together with the presence or absence of a particular property (i.e. a label of the vertex), we can model the similarity measures as the similarity of the set of actions.

\textbf{Action-based definitions}. In this setting, which is gaining increasing attention in the literature, entities can be grouped by the set of actions they perform inside the network. For example, in \cite{tangkdd08} a multi-mode network is considered in which users are connected to queries and ads. Two users are seen as being part of the same community if they are connected to the same queries (i.e. they perform the same actions) even if they are not directly linked to each other. The discovery of communities based on this definition can be performed considering or not the presence of a direct link between entities. Both cases are included in our meta definition.

\textbf{Influence Propagation-based definitions}. In some works, the concept of a ``tribe'' has been introduced. In \cite{gurumine}, a tribe is defined as a set of entities that are influenced by the same leaders. A node is a leader if it has performed an action and, within a chosen time bound after this action, a sufficient number of other users have performed the same action. The role of social ties in this influence spread is considered. Thus, according to our definition, the set of users that frequently perform the same actions due to the influence of their leaders are considered as being a community.

\subsection{Problem Features}

\begin{figure}
\centering
\subfigure[Overlapping Communities]{\includegraphics{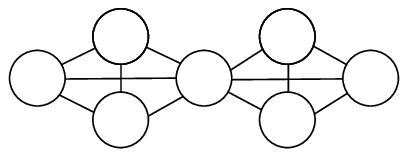}\label{fig:overlapsample}}
\subfigure[Directed Community]{\includegraphics{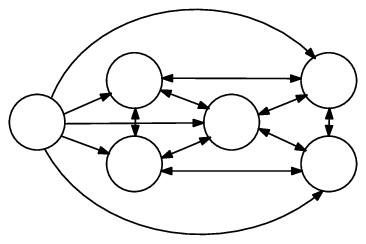}\label{fig:directsample}}
\subfigure[Weighted Communities]{\includegraphics{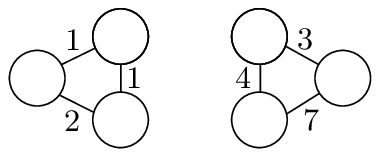}\label{fig:weightsample}}
\caption{Different community features.} 
\end{figure}

There are many features to be considered in the complex task of detecting communities in graph structures. In this section we present some of the features an analyst may be interested in for discovery network communities. We will use them to evaluate the reviewed algorithms in Table 2 and also to motivate our classification in Section \ref{sec:organization}.

Table 2 records the main properties of a community discovery algorithm. These properties can be grouped into two classes. The first class considers the features of the problem representation, the second the characteristics of the approach.

Within the first class of features we group together all the possible variants in the representation of the original real world phenomenon. The most important features we consider are:

\begin{itemize}
\item \textbf{Overlapping}. In some real world networks, communities can share one or more common nodes. For example, in social networks actors may be part of different communities: work, family, friends and so on. All these communities will share a common member, and usually more since a work colleague can also be a friend outside the working environment. Figure \ref{fig:overlapsample} shows an example of possible overlapping community partitions: the central node is shared by the two communities. Table 2 indicates if an algorithm considers this feature in the ``Overlap'' column.
\item \textbf{Directed}. Some phenomena in the real world must be represented with edges and links that are not reciprocal. This, for example, is the case of the web graph: a hyperlink from one page to another is directed and the other page may not have another hyperlink pointing in the other direction. Figure  \ref{fig:directsample} shows an example in which the direction of the edges should be considered. The leftmost node is connected to the community, but only in one direction. If reciprocity is an important feature, the leftmost node should be considered outside the depicted community. See ``Dir'' column in Table 2. 
\item \textbf{Weighted}. A group of connected vertices can be considered as a community only if the weights of their connections are strong enough, i.e. over a given threshold. In the case of Figure \ref{fig:weightsample}, the left group might not be strong enough to form a community. See ``Weight'' column in Table 2.
\item \textbf{Dynamic}. Following our problem representation in Section \ref{sec:probrep}, in our setting we have a set of edges that can appear and disappear. Thus, communities might also evolve over time. See ``Dyn'' column in Table 2.
\end{itemize}

The second class of features collects various desired properties that an approach might have. These features can specify constraints for input data, improve the expressive power of the results or facilitate the community discovery task.

\begin{itemize}
\item \textbf{Parameter free}. A desired feature of an algorithm, especially in data mining research, is the absence of parameters. In other words, an algorithm should be able to make explicit the knowledge that is hidden inside the data without needing any further information from the analyst regarding the data or the problem (for instance the number of communities). See ``NoPar'' column in Table 2.
\item \textbf{Multidimensional input}. Multidimensionality in networks is an emerging topic \cite{szell,signed-networks,asonam-foundations}. A network is said to be multidimensional if it contains a number of different kinds of relations that are established between the nodes of the network. Thus, when dealing with multiple dimensions, the notion of community changes. Our proposed Meta Definition \ref{def:community} captures this complex environment by representing the creation or the absence of a particular edge in a particular dimension with an action. This concept of multidimensionality is used (with various names: multi-relational, multiplex, and so on) by some approaches as a feature of the input considered by the approach. See ``MDim'' column in Table 2.
\item \textbf{Incremental}. Another desired feature of an algorithm is its ability to provide an output without an exhaustive search of the entire input. An incremental approach to the community discovery is to classify a node in one community by looking only at its neighborhood, or the set of nodes two hops away. Alternatively newcomers are put in one of the previously defined communities without starting the community detection process from the beginning. See ``Incr'' column in Table 2.
\item \textbf{Multipartite input}. Many community discovery approaches work even if the network has the particular form of a multipartite graph. The multipartite graph, however, is not entirely a feature of the input that we might want to consider for the output. Many algorithms often use a (usually) bipartite projection of a classical graph in order to apply efficient computations. As in the case of multidimensionality, this is the reason for including the multipartite input as a feature of the approach and not of the output. See ``Multip'' column in Table 2. 
\end{itemize}

There is one more ``meta feature'' that we consider. This is the possibility of applying the considered approach to another community discovery technique by adding new features to the ``guest method''. This meta feature will be highlighted with an asterisk next to the algorithm's name.

Table 2 also has a ``Complexity'' column that gives the time complexity of the methods presented. The two ``BES'' columns give the Biggest Experiment Size, in terms of nodes (``BESn'') and edges (``BESm''), that are included in the original paper reviewed. Note that the Complexity and BES columns often offer an evaluation of the actual values, since the original work did not provide an explicit and clear analysis of the complexity or their experimental setting. A question mark indicates where evaluating the complexity would not be straightforward, or where no experimental details are provided.

\section{The Definition-based classification}\label{sec:organization}
We now review community detection approaches. In each section we group together all the algorithms that share the same definition of what a community is, i.e. the same conditions satisfied by a group of entities that allow them to be clustered together in a community.

This classification is the main contribution of the paper and it should help to get a higher level view of the universe of graph clustering algorithms, by uncovering a practical and reasoned point of view for those analysts seeking to obtain precise results in their analytical problems.

The proposed categories are the following:

\begin{itemize}
\item \textbf{Feature Distance} (Section \ref{sec:firstcategory}). Here we collect all the community discovery approaches that start from the assumption that a community is composed of entities which ubiquitously share a very precise set of features, with similar values (i.e. defining a distance measure on their features, the entities are all close to each other). A common feature can be an edge or any attribute linked to the entity (in our problem definition: the action). Usually, these approaches propose this community definition in order to apply classical data mining clustering techniques, such as the Minimum Description Length principle \cite{mdl, mdl-book}.
\item \textbf{Internal Density} (Section \ref{sec:modularity}). In this group we consider the most important articles that define community discovery as a process driven by directly detecting the denser areas of the network. 
\item \textbf{Bridge Detection} (Section \ref{sec:bridgecommunity}). This section includes the community discovery approaches based on the concept that communities are dense parts of the graph among which there are very few edges that can break the network down into pieces if they are removed. These edges are ``bridges'' and the components of the network resulting from their removal are the desired communities. 
\item \textbf{Diffusion} (Section \ref{sec:percolation}). Here we include all the approaches to the community discovery task that rely on the idea that communities are groups of nodes that can be influenced by the diffusion of a certain property or information inside the network. In addition, the community definition can be narrowed down to the groups that are only influenced by the very same set of diffusion sources. 
\item \textbf{Closeness} (Section \ref{sec:closeness}). A community can also be defined as a group of entities that can reach each of its own community companions with very few hops on the edges of the graph, while the entities outside the community are significantly farther apart.
\item \textbf{Structure} (Section \ref{sec:structure}). Another approach to community discovery is to define the community exactly as a very precise and almost immutable structure of edges. Often these structures are defined as a combination of smaller network motifs. The algorithms following this approach define some kinds of structures and then try to find them efficiently inside the graph. 
\item \textbf{Link Clustering} (Section \ref{sec:linkcommunity}). This class can be viewed as a projection of the community discovery problem. Instead of clustering the nodes of a network, these approaches state that it is the relation that belongs to a community, not the node. Therefore they cluster the edges of the network and thus the nodes belong to the set of communities of their edges. 
\item \textbf{No Definition} (Section \ref{sec:lastcategory}). There are a number of community discovery frameworks which do not have a basic definition of the characteristic of the community they want to explore. Instead they define various operations and algorithms to combine the results of various community discovery approaches and then use the target method community definition for their results. Alternatively, they let the analyst define his / her own notion of community and search for it in the graph.
\end{itemize}

In each section we clarify which features in a particular community discovery category of the ones presented in the previous section are derived naturally, and which features are naturally difficult to achieve. We are not formally building an axiomatic approach, such as the one built in \cite{kleinberg-spatial} for spatial clustering. Instead, we are using the features presented and an experimental setting to make the rationale and the properties of each category in this classification more explicit. The experiments made to support this point are presented in Section \ref{sec:experiments}.

Where possible, we also provide a simple graphical example of the definition considered. This graphical example will provide the main properties of the given classification, in terms of the strong and weak points in particular community features.

The aim of this survey is to focus on the most recent approaches and on the more general definitions of community. We will not focus on historical approaches. Some examples of classical clustering algorithms that have not been extensively reviewed are the Kernighan-Lin algorithm \cite{kernighanlin} or the classical spectral bisection approach \cite{classical-spectral}. Thus, for a historical point of view of the community discovery problem, please refer to other review papers.

\subsection{The Classification Overlap}\label{sec:classoverlap}
There is a sort of overlap for some community definitions. For example a definition of internal density may also include communities with sparse external links, i.e. bridges. We will see in Section \ref{sec:modularity} that in this definition a key concept is modularity \cite{clauset-modularity}. Modularity is a quality function which considers both the internal density of a community and the absence of edges between communities. Thus methods based on modularity could be clustered in both categories. However, the underlying definition of modularity focuses on the internal density, which is the reason for the proposed classification. To give another example, a diffusion approach may detect the same communities whose members can reach each other with just a few hops. However this is not always the case: the diffusion approach may also find communities with an arbitrary distance between its members.

Many approaches in the literature do not explicitly define the communities they want to detect or, worse, they generically claim that their aim is to find dense modules of the network. This is not a problem for us, since the underlying community definition can be inferred from a high-level understanding of the approach described in the original paper. One cannot expect researchers to be able to categorize their method before an established categorization has been accepted. To instigate a discussion regarding this issue is one of the aims of this paper. Once further knowledge regarding the field has been established, authors will be able to correctly categorize their approach.

In order to gain stronger evidence of the differences between the proposed categories, consider Figures \ref{fig:metadef-featdist}, \ref{fig:metadef-intdensity}, \ref{fig:metadef-bridge}, \ref{fig:metadef-diffusion}, \ref{fig:metadef-closeness} and \ref{fig:cliqueperc}. These figures depict the simplest typical communities that have been identified from the definitions of Feature Distance, Internal Density, Bride Detection, Diffusion, Closeness and Structure Definition, respectively. As can be seen, there are a number of differences between these examples. The Bridge Detection example (Figure \ref{fig:metadef-bridge}) is a random graph, thus with no community structure defined for the algorithms in the Internal Density category. The Diffusion example (Figure \ref{fig:metadef-diffusion}) is also a random graph, however although the diffusion process identifies two communities, no clear bridges can be detected.

The overlap is due to the fact that many algorithms work with some general ``background'' meta definitions of community. The categories proposed here can be clustered together into a hierarchy with the four main categories described in Section \ref{sec:metadef}. Further, many algorithms may present common strategies in the exploration of the search space or in evaluating the quality of their partition in order to refine it. Consider for example \cite{lambiotte-laplacian} and \cite{reichardt}. In these two papers there is a thorough theoretical study concerning modularity and its most general form. In \cite{lambiotte-laplacian}, for example, the authors were able to derive modularity as a random walk exploration strategy, thus highlighting its overlap with the algorithms clustered here in the ``Closeness'' category.

Evaluating the overlap and the relationships between the most important community discovery approaches is not simple, and is outside the scope of this survey. Here we focus on the connection between an algorithm and its particular definition of community. Thus we can create our useful high-level classification to connect the needs of particular analyses (i.e. the community definitions) to the tools available in the literature. To study how to derive one algorithm in terms of another, thus creating a graph of algorithms and not a classification, is an interesting open issue which we will leave for future research.

\begin{table*}\vspace{-1.5cm}
\begin{center}
\begin{turn}{270}
\begin{tabular}{|c|c|cccccccc|ccc|c|c|}
\hline
 & Name & Overlap & Dir & Weight & Dyn & NoPar & MDim & Incr & Multip & Complexity & BESn & BESm & Year & Ref \\
\hline
\multirow{10}{*}{\begin{sideways}Feature Distance\end{sideways}}
& Evolutionary* & & & & $ \checkmark $ & & & $ \checkmark $ & & $\mathcal{O}(n^{2})$ & 5k & ? & 2006 & \cite{evolutionary-clustering}\\
& MSN-BD & & & $ \checkmark $ & & & & & $ \checkmark $ & $\mathcal{O}(n^{2}ck)$ & 6k & 3M & 2006 & \cite{clustering-kpartite} \\
& SocDim & $ \checkmark $ & & $ \checkmark $ & & & $ \checkmark $ & & & $\mathcal{O}(n^{2} \log n)*$ & 80k & 6M & 2009 & \cite{tangkdd} \\
& PMM & & & $ \checkmark $ & & & $ \checkmark $ & & & $\mathcal{O}(mn^{2})$ & 15k & 27M & 2009 & \cite{tangicdm} \\
& MRGC & & $ \checkmark $ & & $ \checkmark $ & & $ \checkmark $ & & $ \checkmark $ & $\mathcal{O}(mD)$ & 40k & ? & 2007 & \cite{tensor-clustering}\\
& Infinite Relational & & & & & $ \checkmark $ & $ \checkmark $ & & & $\mathcal{O}(n^{2c}D)$ & 160 & ? & 2006 & \cite{irm}\\
& Find-Tribes & & & & $ \checkmark $ & & & & $ \checkmark $ & $\mathcal{O}(mnK^{2})$ & 26k & 100k & 2007 & \cite{findtribes}\\
& AutoPart & & $ \checkmark $ & & & $ \checkmark $ & & $ \checkmark $ & & $\mathcal{O}(mk^{2})$ & 75k & 500k & 2004 & \cite{autopart} \\
& Timefall & & & & $ \checkmark $ & $ \checkmark $ & & & $ \checkmark $ & $\mathcal{O}(mk)$ & 7.5M & 53M & 2008 & \cite{timefall} \\
& Context-specific Cluster Tree & & & & & $ \checkmark $ & & & $ \checkmark $ & $\mathcal{O}(mk)$ & 37k & 367k & 2008 & \cite{cct} \\
\hline
\multirow{5}{*}{\begin{sideways}IntDensity\end{sideways}}
& Modularity & $ \checkmark $ & $ \checkmark $ & $ \checkmark $ & & & $ \checkmark $ & $ \checkmark $ & $ \checkmark $ & $\mathcal{O}(mk \log n)$ & 118M & 1B & 2004 & \cite{clauset-modularity} \\
& MetaFac & & & & $ \checkmark $ & & $ \checkmark $ & & & $\mathcal{O}(mnD)$ & ? & 2M & 2009 & \cite{metafac}\\
& Variational Bayes & & $ \checkmark $ & & & $ \checkmark $ & & & & $\mathcal{O}(mk)$ & 115 & 613 & 2008 & \cite{hofman-bayesian}\\
& $LA \rightarrow IS^{2}$* & $ \checkmark $ & $ \checkmark $ & & & & & & & $\mathcal{O}(mk + n)$ & 16k & ? & 2005 & \cite{densityoverlap}\\
& Local Density & & $ \checkmark $ & & & $ \checkmark $ & & $ \checkmark $ & & $\mathcal{O}(nK \log n)$ & 108k & 330k & 2005 & \cite{localdensity}\\
\hline
\multirow{4}{*}{\begin{sideways}Bridge\end{sideways}}
& Edge Betweenness & & $ \checkmark $ & $ \checkmark $ & & & & & & $\mathcal{O}(m^{2}n)$ & 271 & 1k & 2002 & \cite{edgebetween}\\
& CONGO* & $ \checkmark $ & & $ \checkmark $ & & & & & & $\mathcal{O}(n \log n)$ & 30k & 116k & 2008 & \cite{conga2} \\
& L-Shell & $ \checkmark $ & & & & & & $ \checkmark $ & & $\mathcal{O}(n^{3})$ & 77 & 254 & 2005 & \cite{lshell}\\
& Internal-External Degree & $ \checkmark $ & & & & & & & & $\mathcal{O}(n^{2} \log n)$ & 775k & 4.7M & 2009 & \cite{lancichinetti}\\
\hline
\multirow{7}{*}{\begin{sideways}Diffusion\end{sideways}}
& Label Propagation & & & $ \checkmark $ & & $ \checkmark $ & & $ \checkmark $ & & $\mathcal{O}(m + n)$ & 374k & 30M & 2007 & \cite{labelprop}\\
& Node Colouring & & & & $ \checkmark $ & & & & $ \checkmark $ & $\mathcal{O}(ntk^{2})$ & 2k & ? & 2007 & \cite{node-coloring}\\
& Kirchhoff & $ \checkmark $ & & $ \checkmark $ & & & & & & $\mathcal{O}(m + n)$ & 115 & 613 & 2004 & \cite{kirchhoff}\\
& Communication Dynamic & $ \checkmark $ & $ \checkmark $ & & $ \checkmark $ & & & $ \checkmark $ & & $\mathcal{O}(mnt)$ & 160k & 530k & 2008 & \cite{commdyn}\\
& GuruMine & & $ \checkmark $ & & $ \checkmark $ & & & & & $\mathcal{O}(TAn^{2})$ & 217k & 212k & 2008 & \cite{gurumine}\\
& DegreeDiscountIC & & $ \checkmark $ & & & & & & & $\mathcal{O}(k \log n + m)$ & 37k & 230k & 2009 & \cite{flu-maxim}\\
& MMSB & $ \checkmark $ & $ \checkmark $ & & & & & & & $\mathcal{O}(nk)$ & 871 & 2k & 2007 & \cite{mixedmembership}\\
\hline
\multirow{3}{*}{\begin{sideways}Close\end{sideways}}
& Walktrap & & & $ \checkmark $ & & & & & & $\mathcal{O}(mn^{2})$ & 160k & 1.8M & 2006 & \cite{walktrap}\\
& DOCS & $ \checkmark $ & & & & & & & & ? & 325k & 1M & 2009 & \cite{docs}\\
& Infomap & & $ \checkmark $ & $ \checkmark $ & & $ \checkmark $ & & & & $\mathcal{O}(m \log ^{2}n)$ & 6k & 6M & 2008 & \cite{infomap}\\
\hline
\multirow{4}{*}{\begin{sideways}Structure\end{sideways}}
& K-Clique & $ \checkmark $ & & & & & & & & $\mathcal{O}(m^{\frac{\ln m}{10}})$ & 20k & 127k & 2005 & \cite{kclique}\\
& S-Plexes Enumeration & $ \checkmark $ & & & & & & & & $\mathcal{O}(kmn)$ & ? & ? & 2009 & \cite{splexes}\\
& Bi-Clique & $ \checkmark $ & & & & & & & $ \checkmark $ & $\mathcal{O}(m^{2})$ & 200k & 500k & 2008 & \cite{biclique}\\
& EAGLE & $ \checkmark $ & $ \checkmark $ & $ \checkmark $ & & & & & & $\mathcal{O}(3^{\frac{n}{3}})$ & 16k & 31k & 2009 & \cite{eagle}\\
\hline
\multirow{3}{*}{\begin{sideways}Link\end{sideways}}
& Link modularity & $ \checkmark $ & & $ \checkmark $ & & & & & $ \checkmark $ & $\mathcal{O}(2mk \log n)$ & 20k & 127k & 2009 & \cite{link-modularity} \\
& HLC* & $ \checkmark $ & & $ \checkmark $ & & & & & $ \checkmark $ & $\mathcal{O}(n\bar{K}^{2})$ & 885k & 5.5M & 2010 & \cite{link-jaccard}\\
& Link Maximum Likelihood & $ \checkmark $ & & & & & & & & $\mathcal{O}(mk)$ & 4.8M & 42M & 2011 & \cite{link-bayes}\\
\hline
\multirow{4}{*}{\begin{sideways}NoD\end{sideways}}
& Hybrid* & $ \checkmark $ & $ \checkmark $ & $ \checkmark $ & & $ \checkmark $ & & & & $\mathcal{O}(nk\bar{K})$ & 325k & 1.5M & 2010 & \cite{bayeshybrid}\\
& Multi-relational Regression & & & $ \checkmark $ & & & $ \checkmark $ & & & ? & ? & ? & 2005 & \cite{multirelationalhan}\\
& Hierarchical Bayes & & & & & & & & & $\mathcal{O}(n^2)$ & 1k & 4k & 2008 & \cite{clauset-bay-hier}\\
& Expectation Maximization & & $ \checkmark $ & & & & & & & ? & 112 & ? & 2007 & \cite{leicht-bayes-def}\\
\hline
\end{tabular}
\end{turn}
\end{center}
\label{tab:algorithmresume}
\caption{Resume of the community discovery methods.}
\end{table*}

\section{Feature Distance}\label{sec:firstcategory}

In this section we review the community discovery methods that define a community according to this meta definition:

\begin{definition}[Feature Community]\label{def:featdistcommunity}
A feature community in a complex network is a set of entities that share a precise set of features (including the edge as a feature). Defining a distance measure based on the values of the features, the entities inside a community are very close to each other, more than the entities outside the community. 
\end{definition}

This meta definition operates according to the following meta procedure:

\begin{procedure}Given a set of entities and their attributes (which may be relations, actions or properties), represent them as a vector of values according to these attributes and thus operate a matrix/spatial clustering on the resulting structure. \end{procedure}

Using this definition the task of finding communities is very similar to the classical clustering problem in data mining. In data mining, clustering is an unsupervised learning task. The aim of a clustering algorithm is to assign a large set of data into groups (clusters) so that the data in the same clusters are more similar to each other than any other data in any other cluster. Similarity is defined through a distance measure, usually based on the number of common features of the entities, or on similar values of these attributes.

An example of the clustering technique is K-means \cite{kaufman-means}. One natural clustering approach to the community discovery is some evolutions of co-clustering \cite{coclustering, cct-base} and/or some spectral approaches to the clustering problem \cite{spectral-clustering}. In \cite{biclustering-survey} there is a survey on co-clustering algorithms, while in \cite{kleinberg-spatial} there is an interesting axiomatic framework for spatial clustering. Given the rich literature and methods to cluster matrices, in this category community discovery approaches may find clusters with virtually any feature we presented. Table 2 illustrates this by looking at the high entropy of the features set for all methods present in this category. Given the fact that each node and edge is represented by a set of attributes, it is very easy to obtain multidimensional and multi-partite results by simply clustering it in a complex multidimensional space.

\begin{figure}
\centering
\includegraphics[scale=0.6]{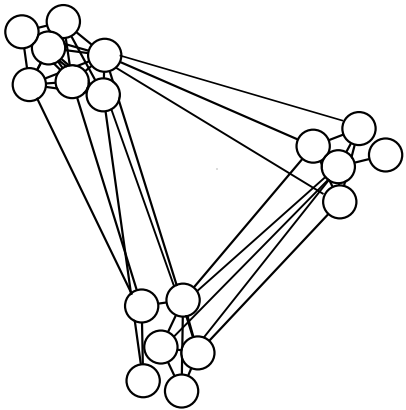}
\caption{An example of a graph that can be partitioned with a notion of ``distance'' between its nodes.}\label{fig:metadef-featdist}
\end{figure}

In order to understand the downsides of this category, consider Figure \ref{fig:metadef-featdist}, which depicts a network whose nodes are positioned according to a distance measure. This measure could consider the direct edge connection, however it is not mandatory. The nodes are then grouped into the same community if they are close in this space (which may be highly dimensional depending on the number of features considered). Figure \ref{fig:metadef-featdist} shows that, depending on the number of node/edge attributes, the underlying graph structure may lose importance. This may lead to counter-intuitive results if the analyst tries to display the clusters by only looking at the graph structure, thus resulting in a lot of inter-community edges. We will discuss this point further in Section \ref{sec:experiments}.

Here we focus on some clustering techniques with some very interesting features: the Evolutionary clustering \cite{evolutionary-clustering}; RSN-BD \cite{clustering-kpartite}, a k-partite graph based approach; MRGC \cite{tensor-clustering}, that is a clustering technique working with tensors; two approaches that use modularity for the detection of latent dimensions for a multidimensional community discovery with a machine learning classifier that maximizes the number of common features (\cite{tangkdd} and \cite{tangicdm}); a Bayesian approach to clustering based on the predictability of the features for nodes belonging to the same group \cite{irm}; and an analysis of the shared attribute connections in a bipartite graph entity-attribute \cite{findtribes}.

An interesting clustering principle is the Minimum Description Length principle \cite{mdl, mdl-book}. In MDL the main concept is that any regularity in the data (i.e. common features) can be used to compress it, i.e. to describe it using fewer symbols than the number of symbols needed to describe the data literally (see also \cite{infotheory} and \cite{mdl-tutorial}). The more regularities there are, the more the data can be compressed. This is a very interesting approach since, in some implementations, it enables the community discovery to be performed without setting any parameters. After considering the classical clustering approaches, in this section we also present three main algorithms that implement a MDL community discovery approach: Autopart \cite{autopart} (that is, to the best of our knowledge, the first popular community discovery that formulates the ground theory for the MDL community detection), the Context-specific cluster tree \cite{cct}, and Timefall \cite{timefall}.

\subsection{Evolutionary* \cite{evolutionary-clustering}}
In \cite{evolutionary-clustering} the authors tackle the classical clustering problem by adding a temporal dimension. This novel situation includes several constraints:

\begin{itemize}
\item \textbf{Consistency}. Any insights derived from a study of previous clusters are more likely to apply to future clusters. 
\item \textbf{Noise Removal}. Historically consistent clustering provides greater robustness against noise by taking previous data points into effect. 
\item \textbf{Smoothing}. The true clusters shift over time. 
\item \textbf{Cluster Correspondence}. It is generally possible to place today's clusters in relation to yesterday's clusters, so the user will still be situated within the historical context. 
\end{itemize}

In order to consider these constraints, two clustering division measures are defined: snapshot quality and history cost. The snapshot quality of $C_{t}$, a proposed cluster division, measures how well $C_{t}$ represents the data at time-step $t$. The history cost of the clustering is a measure of the distance between $C_{t}$ and $C_{t-1}$, the clustering used during the previous time-step.

This setting is similar to incremental clustering, but with some differences, \cite{incremental-clustering}. There are two main differences. First, the focus is on optimizing a new quality measure which incorporates a deviation from history. Secondly, it works on-line (i.e. it must cluster the data during time-step $t$ before seeing any data for time-step $t+1$), while other frameworks work on data streams \cite{stream-clustering}.

This framework can be added to any clustering algorithm. The time complexity will be $\mathcal{O}(n^{2})$, particularly on the agglomerative hierarchical clustering, used for the examples in the original paper, although some authors claim that a quasi-linear implementation \cite{linear-hier} is possible. However, the framework is presented here because it is possible to apply its principles to all the other community discovery algorithms presented in this survey. 

There are two framework applications worth noting. The first is FacetNet \cite{facetnet}, in which a framework to evaluate the evolution of the communities is developed. The second one is \cite{kll}, in which the concepts of nano-communities and k-clique-by-clique are introduced. These concepts are useful for assessing the snapshots and historical quality of the communities identified in various snapshots with any given method.

\subsection{RSN-BD \cite{clustering-kpartite}}
RSN-BD (Relation Summary Network with Bregman Divergence) is a community discovery approach focused on examples of real-world data that involve multiple types of objects that are related to each other. A natural representation of this setting is a k-partite graph of heterogeneous types of nodes. This method is suitable for general k-partite graphs and not only special cases such as \cite{old-kpartite}. The latter has the restriction that the numbers of clusters for different types of nodes must be equal, and the clusters for different types of objects must have one-to-one associations.

The key idea is that in a sparse k-partite graph, two nodes are similar when they are connected to similar nodes even though they are not connected to the same nodes. In order to spot this similarity, authors produce a derived structure (i.e. a projection) to make these two node closely connected. In order to do this, the authors of \cite{clustering-kpartite} add a small number of hidden nodes. This derived structure is called a Relation Summary Network and must be as close as possible to the original graph. They can evaluate the distance between the two structures by linking every original node with one hidden node and every hidden node couple if both hidden nodes are linked by the same original node. The distance function then sums up all the Euclidean distances between the weights of the edges in the original graph and in the transformed graph (any Bregman divergence distance function can be used). A Bregman divergence defines a class of distance measures for which neither the triangle inequality, nor symmetry, is respected, and these measures are defined for matrices, functions and distributions \cite{distance-functions}. The total complexity of the algorithm, as discussed by the authors, is $\mathcal{O}(n^2ck)$.

\subsection{MRGC \cite{tensor-clustering}}\label{sec:mrgc}
In this model, each relation between a given set of entity classes is represented as a multidimensional tensor (or data cube) over an appropriate domain, with the dimensions associated with the various entity classes. In addition, each cell in the tensor encodes the relation between a particular set of entities and can either take real values, i.e., the relation has a single attribute, or itself is a vector of attributes.

The general idea is that each node and each relation is a collection of attributes. All these attributes are a dimension of the relational space. MRGC (Multi-way Relation Graphs Clustering), basically tries to find a solution on one dimension at a time. It finds the optimal clustering with respect to each dimension by keeping every other intermediate result on the other dimensions fixed (thus its time complexity is given by the number of relations times the number of dimensions, i.e. $\mathcal{O}(mD)$). It then evaluates the solutions and keeps recalculating over all dimensions until it converges. Although defined for relation graphs, this model can be also used for identify community structures in social networks.

MRGC operates in a multi-way clustering setting where the objective is to map the set of entities in a (smaller) set of clusters by using a set of clustering functions (i.e. it is a general framework in which previous co-clustering approaches, such as \cite{cocluster2}, can be viewed as special cases). The crucial mechanism in this problem is how to evaluate the quality of the multi-way clustering in order to get to the convergence. In this case, the authors propose to measure it in terms of the approximation error or the expected Bregman distortion \cite{bregman} between the original tensor and the approximate tensor built after applying the clustering function.

\subsection{SocDim \cite{tangkdd}}
One basic (Markov) assumption in community discovery is frequently that the label of a node is only dependent on the labels of all its neighbors. SocDim tries to go beyond this assumption by building a classifier which not only considers the connectivity of a node, but assigns additional information to its connection i.e. a description of a likely affiliation between social actors. This information is called latent social dimensions and the resulting framework is based on relational learning.

In order to do this, two steps are performed by SocDim. Firstly, it extracts latent social dimensions based on network connectivity. It uses modularity (Section \ref{sec:modularity}) in order to find in the structure of the network the dimensions in which the nodes are placed (following the homophily theory which states that actors sharing certain properties tend to form groups \cite{homophily}). This can usually be done in $\mathcal{O}(n^2\log n)$. This step may be replaced if there is already knowledge of the social dimensions. Secondly, it constructs a discriminative classifier (one-vs-rest linear \cite{metalabeler} or structural \cite{svm} SVM): the extracted social dimensions are considered as normal features (including other possible sources) in the classical supervised learning task. It is then possible to use the predicted labels of the classifier to reconstruct the community organization of the entities. This is a multidimensional community discovery because the classifier will determine which dimensions are relevant to a class label. 

This work is the basis of a further evolution \cite{tangcikm} that has an edge-centric view of communities (similar to the methods classified in Section \ref{sec:linkcommunity})

\subsection{PMM \cite{tangicdm}}
This work was originally presented in \cite{tangsdm} and then evolved in \cite{tangicdm}. It presents a variation of the modularity approach on a multidimensional setting. The goal of the PMM (Principal modularity Maximization) algorithm is: given a lot of different dimensions, find a concise representation of them (the authors call this step ``Structural Feature Extraction'', computing modularity with the Lanczos method. The latter is an algorithm to find eigenvalues and eigenvectors of a square matrix \cite{matrix}, of complexity $\mathcal{O}(mn^{2})$) and then detect the correlations between these representations (in the ``Cross-Dimension Integration'', using a generalized canonical correlation analysis \cite{cca}).

After this step, the authors obtain lower-dimensional embedding, which captures the principal pattern across all the dimensions of the network. They can then perform k-means \cite{kaufman-means} on this embedding to find out the discrete community assignment.

\subsection{Infinite Relational \cite{irm}}
Suppose there are one or more relations (i.e. edges) involving one or more types (i.e. nodes). The goal of the Infinite Relational Model is to partition each type into clusters (i.e. communities), where a good set of partitions allows relationships between entities to be predicted by their cluster assignments. The authors' goal is to organize the entities into clusters that relate to each other in predictable ways, by simultaneously clustering the entities and the relations.

Formally, suppose that the observed data are $m$ relations involving $n$ types. Let $R^i$ be the $i$th relation, $T^j$ be the $j$th type, and $z^j$ be a vector of cluster assignments for $T^j$. The task is to infer the cluster assignments, and the ultimate interest lies in the posterior distribution $P(z_1, ... , z_n \mid R_1, ... , R_m)$.

To enable the IRM to discover the number of clusters in type $T$, the authors use a prior \cite{pitmancombinatorial} that assigns some probability mass to all possible partitions of the type. Inferences can be made using Markov chain Monte Carlo methods to sample from the posterior on cluster assignments. This method has a very high time complexity ($\mathcal{O}(n^{2c}D)$).

\subsection{Find-Tribes \cite{findtribes}}
Find-Tribes was not explicitly developed for community discovery purposes. However, the technique can still be used to identify some kind of community. It is very close to our ``action'' definition of a community: the entities in a group tend to behave in the same way.

As input, the authors require a bipartite graph $G = (R \cup A, E)$ of entities $R$ and attributes $A$. The entities should connect to several attributes. The aim of the algorithm is to return those groups sharing ``unusual'' combinations of attributes. This restriction can be easily generalized in order to also obtain the ``usual'' groups as outputs.

The strategy for the desired task revolves around the development of a good definition of ``unusual''. For an entity group to be considered anomalous, the shared attributes themselves need not be unusual, but their particular configuration should be. A projected non-bipartite graph $H'(R,F)$ is built, then for each edge a score $c_{ij}$ (the number of attributes in the shared sequence, the number of time steps of overlap, a probabilistic Markov chain of attributes and so on) is computed, measuring how significant or unusual its sequence of shared attributes is. In the end a threshold $d$ is chosen and all edges $f_{ij}$ removed for which $c_{ij} < d$ are removed. The connected components of $H'$ are the desired tribes and the overall complexity is $\mathcal{O}(mnK^2)$.

\begin{figure}
\centering
\subfigure[The original matrix]{\includegraphics[scale=0.3]{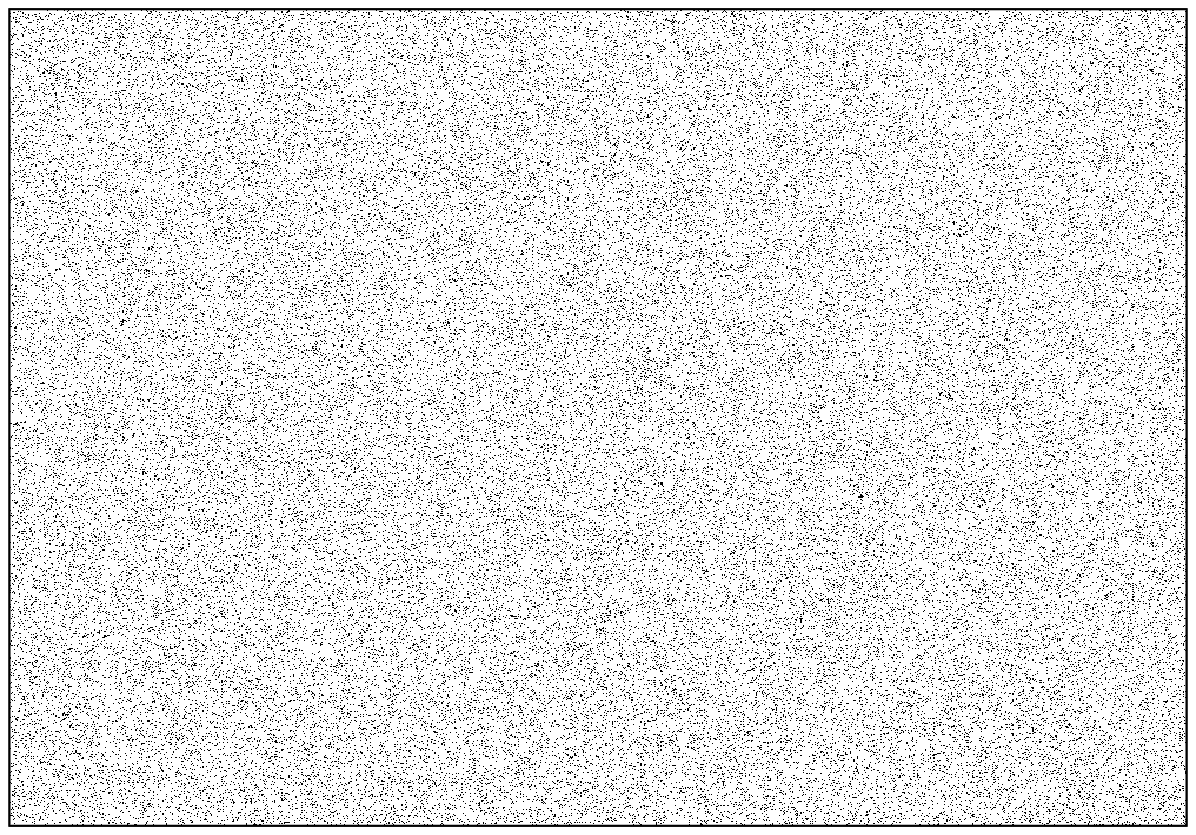}\label{fig:rndmatrix}}
\subfigure[Reordered matrix]{\includegraphics[scale=0.3]{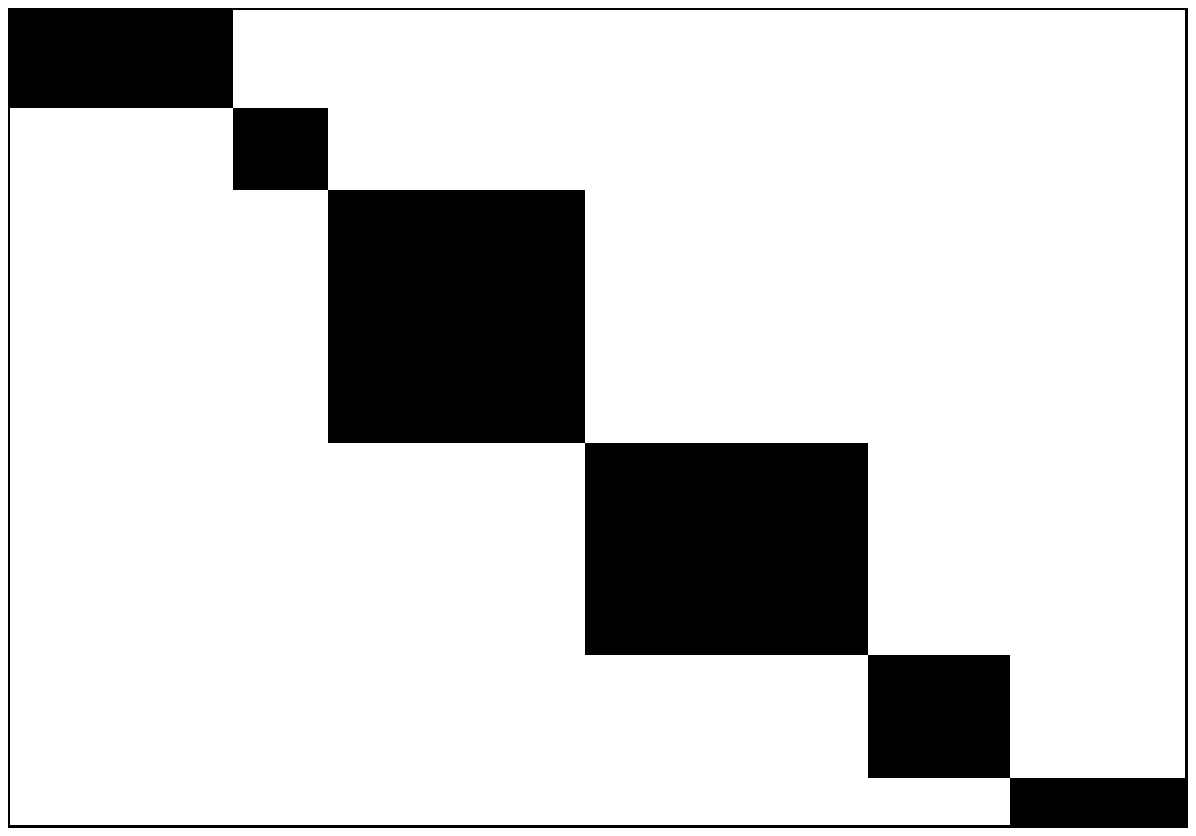}\label{fig:orderedmatrix}}
\caption{An example of the MDL principle for matrices: the matrix on the left is exactly the same matrix as the one on the right, but reordered in order to describe it simply.} 
\end{figure}

\subsection{AutoPart \cite{autopart}}
Autopart is the basic formulation of the MDL approach to the community discovery problem. There is a binary matrix that represents associations between the $n$ nodes of the graph (and their attributes). An example of a possible adjacency matrix is shown in Figure \ref{fig:rndmatrix}.

The main idea is to reorder the adjacency matrix so that similar nodes, i.e. nodes that are connected to the same set of nodes, are grouped with each other. The adjacency matrix should then consist of homogeneous rectangular/square blocks of a high (low) density, representing the fact that certain node groups have more (less) connections with other groups (right hand side of Figure \ref{fig:orderedmatrix}), which can be encoded with a great compression of the data. The aim of the algorithm is to identify the best grouping that minimizes the cost (compression) function \cite{mdl-parameter-free}.

A trade-off point must therefore be identified that indicates the best number of groups $k$. The authors solved this problem using a two-step iterative process: first, they find a good node grouping $G$ for a given number of node groups $k$ that minimize entropy; and second, they search for the number of node groups $k$ by splitting the previously identified groups and verifying if there is a possible gain in the total encoding cost function, at a total time complexity of $\mathcal{O}(mk^2)$.

\subsection{Context-specific Cluster Tree \cite{cct}}
In this variant of the MDL approach, a binary $n_{s} \times n_{d}$ matrix represents a bipartite graph with $n_{s}$ source nodes and $n_{d}$ destination nodes. The aim is to automatically construct a recursive community structure of a large bipartite graph at multiple levels, namely, a Context-specific Cluster Tree (CCT). The resulting CCT can identify relevant context-specific clusters. The main idea is to subdivide the adjacency matrix into tiles, or ``contexts'', with a possible reordering of rows and columns, and to compress them, either as-is (if they are homogeneous enough) or by further subdividing.

The entire graph is considered as a whole community. If the best representation of the considered (sub)graph is the random graph, by testing its possible compression with a total encoding cost function, then the community cannot be split into two sub-communities. In fact, by definition the random graph has no community structure at all. Otherwise, the graph is split and the algorithm is reapplied recursively. Each edge is visited once for each subdivision (thus the complexity is $\mathcal{O}(mk)$). The result is a tree of communities in which the bottom levels are a context specialization of the generic communities at the top of the tree. 

This idea of recursive clustering is also applied to streaming setting \cite{streaming-biclustering, graphscope}, although with a number of parameters. This is a hierarchical evolution of the existing flat method described in \cite{cct-base}.

\subsection{Timefall \cite{timefall}}
Timefall is an MDL approach that can be described as a parameter-free network evolution tracking. Given $n$ time-stamped events each related to several of $m$ items, it simultaneously finds (a) the communities, that is, item-groups (e.g., research topics and/or research communities) and (b) a description of how the communities evolve over time (e.g., appear, disappear, split, merge), and (c) a selection of the appropriate cut-points in time when existing community structures change abruptly.

The adjacency matrix representing the graph is split according to the row timestamps. Columns are then clustered with a Cross Association algorithm  \cite{cct-base}, which is the basis of the MDL community discovery algorithms. The MDL principle is used again to connect the column clusters of the matrices across the split rows: if two column clusters can be encoded together with a low encoding cost then they are connected, ignoring time points with little or no changes. The time complexity is equal to $\mathcal{O}(mk)$.

\section{Internal Density}\label{sec:modularity}

\begin{figure}
\centering
\includegraphics[scale=0.6]{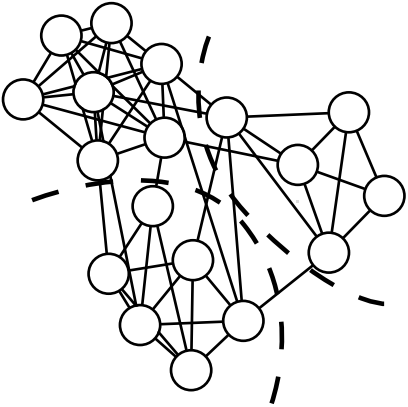}
\caption{An example of a graph which can be partitioned with a notion of internal density between its nodes.}\label{fig:metadef-intdensity}
\end{figure}

For this group of approaches, the underlying meta definition is: 

\begin{definition}[Dense Community]\label{def:intdensitycommunity}
A dense community in a complex network is a set of entities that are densely connected. In order to be densely connected, a group of vertices must have a number of edges significantly higher than the expected number of edges in a random graph (which has no community structure). 
\end{definition}

The following meta procedure is generally shared by the algorithms in this category:

\begin{procedure}
Given a graph, try to expand or collapse the node partitions in order to optimize a given density function, stopping when no increment is possible. 
\end{procedure}

Figure \ref{fig:metadef-intdensity} shows a network in which the identified communities are significantly denser than a random graph with the same degree distribution.

A key concept for satisfying this meta definition is modularity \cite{newman-modularity}. Briefly, consider dividing the graph into $c$ non-overlapping communities. Let $c_i$ denote the community membership of vertex $v_i$, $k_i$ represents the degree of vertex $i$. Modularity is like a statistical test in which the null model is a uniform random graph model. In this model one entity connects to others with uniform probability. For two nodes with degree $k_i$ and $k_j$ respectively, the expected number of edges between the two in a uniform random graph model is $\dfrac{k_i k_j}{2m}$, where $m$ is the number of edges in the graph. Modularity measures how far the interaction deviates from a uniform random graph with the same degree distribution. It is defined as: 

$$Q = \frac{1}{2m} \sum_{ij} \left[ A_{ij} - \frac{k_i k_j}{2m} \right] \delta(c_i,c_j), $$

where $\delta(c_i,c_j) = 1$ if $c_i = c_j$ (i.e. the two nodes are in the same community), and 0 otherwise, and $A_{ij}$ is the number of edges between nodes $i$ and $j$. A larger modularity indicates a denser within-group interaction. Note that $Q$ could be negative if the vertices are split into bad clusters. $Q > 0$ indicates that the clustering captures some degree of community structure. Essentially, the aim is to find a community structure such that $Q$ is maximized.

Modularity is involved in the community discovery problem on two levels. Firstly,  it can quantify how good a given network partition is. It gives a result of the quality of the partition even without any knowledge of the actual communities of the network. This is especially suitable for very large networks. On the other hand, modularity is not the perfect solution for evaluating a proposed community partition. It suffers from well known problems, in particular the resolution problem. Modularity fails to identify communities smaller than a scale that depends on the total size of the network and on the degree of interconnectedness of the communities, even in cases where modules are unambiguously defined. Furthermore, with modularity only communities extracted according to the meta definition proposed in this section can be evaluated. Any other kind of definition of communities will result in a not so meaningful evaluation by applying modularity. For an extensive review of the known problems of modularity see \cite{bfsurvey, nearglobaloptimalpeaks}.

The second level of the modularity usage in the graph partitioning task is represented by community discovery algorithms that are based on modularity maximization. These algorithms suffer from the aforementioned problems of the usage of modularity as quality measures. However, modularity maximization is a very prolific field of research, and there are many algorithms relying on heuristics and strategies for finding the best network partition.

We will present the main example of a modularity-based approach, providing references for minor modularity maximization algorithms. A good review of the eigenvector modularity based work is in \cite{modularity-review}).

Modularity is not the only cost function that is able to quantify whether a set of entities is more related than expected and thus can be considered as a community. The other reviewed methods that rely on different techniques, but share the same meta definition of community proposed in this section, are: MetaFac \cite{metafac}, a hypergraph factorization technique; a physical-chemical algorithm using a Bayesian approach \cite{hofman-bayesian}; a local density-based approach called $LA \rightarrow IS^{2}$ \cite{densityoverlap}; and another proposed function used to measure the internal local density of a cluster \cite{localdensity}.

Optimizing a density function is suitable for many graph representations such as directed graphs and weighted graphs. However in addition to modularity problems, there are other weak points. For example, more complex structures are not tractable in this approach such as multidimensional networks. If multiple different qualitative relations are present in a network, how should a consistent value of ``multirelational density'' be computed?  There are some works that scratch the surface of the ambiguity of density in multidimensional networks \cite{multicd-asonam}, however given the current situation none of these approaches can be used in pure multidimensional settings.

\subsection{Modularity \cite{clauset-modularity}}

\begin{figure}
\centering
\includegraphics[scale=0.66]{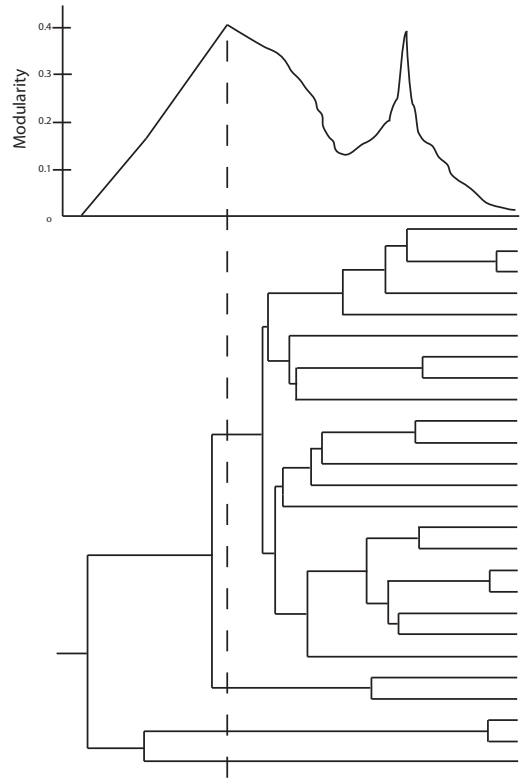}
\caption{A dendrogram result for the modularity maximization algorithm, with a plot of resulting modularity values given the partition.} \label{fig:modmax}
\end{figure}

To find a partition that provides the maximum value of modularity is an NP-complete problem. Many greedy heuristics have therefore been proposed. After a pioneering work proposing modularity \cite{slow-modularity}, Newman presented an efficient strategy for modularity maximization, namely repeatedly merging the two communities whose amalgamation produces the largest increase in $Q$. This produces a dendrogram representing the hierarchical decomposition of the network into communities at all levels, which must be cut in the modularity peak in order to obtain the communities, as depicted in Figure \ref{fig:modmax}.

Figure \ref{fig:modmax} also shows another problem of modularity maximization heuristics. It has been discovered that modularity does not have a single peak given all the possible partitions, but there are several local optima. Moreover, real networks have many near-global-optima at various places \cite{nearglobaloptimalpeaks} (the rightmost peak in Figure \ref{fig:modmax}) and we cannot know where the algorithm locates its solution.

The optimization proposed by Clauset et al. \cite{clauset-modularity} is to store a matrix containing only the values of the communities, i.e. the modularity changes when joining the communities $i$ and $j$. The algorithm can now be defined as follows. Calculate the initial values of $\Delta Q_{i,j}$ and keep track of the largest element of each row of the matrix $\Delta Q$. Select the largest $\Delta Q_{i,j}$ among these largest elements, join the corresponding communities, update the matrix $\Delta Q$ and the collection of the largest elements and increment $Q$ by $\Delta Q_{i,j}$. Repeat this last step until the dendrogram is complete. In \cite{directed-modularity} the modularity maximization approach is adapted to the case of a directed network. We therefore have a matrix representation of the graph, but the matrix is not symmetric. The algorithm is based on \cite{leicht-base}. 

More recent works point to also applying the modularity approach to overlapping communities \cite{nicosia-modularity}. A local evaluation of modularity has also been proposed, by dividing the graph into known, boundary and unexplored sets. Two more implementations of modularity-based algorithms can be found in \cite{mod-impl}.

Another optimization of modularity-based approaches is presented in \cite{external-optimization}. This is basically a divisive algorithm that optimizes the modularity $Q$ using a heuristic search. This search is based on a measure ($\lambda$) that depends on the node degree, and its normalization involves all the links in the network after summation. The node selected, in an original External Optimization algorithm \cite{eo-original} is always the node with the worst $\lambda_{i}$-value. There is a $\tau$-EO version \cite{tau-eo} that is less sensitive to different initializations and allows escape from local maxima. A number of other optimization strategies have been proposed (size reduction \cite{modularity-optim1}, simulated annealing \cite{modularity-optim2}).

Finally, we present the last greedy approach working with the classical definition of modularity \cite{mod-unfolding}. The previous largest graph used for modularity testing was 5.5 million nodes \cite{fast-modularity}, with this improvement it is possible to scale up to 100 million nodes. The algorithm is divided into two phases that are repeated iteratively. For each node $i$ the authors consider the neighbors $J$ of $i$ and evaluate the gain in modularity that would take place by removing $i$ from its community and by placing it in the community of $J$. The node $i$ is then placed in the community for which this gain is maximum until no individual move can improve the modularity. The second phase consists in building a new network whose nodes are now the communities found during the first phase. It is then possible to reapply the first phase to the resulting weighted network and to iterate. This method has been tested on the UK-Union WebGraph \cite{webgraph}, on co-citation networks \cite{blondel-application09}, and on mobile phone networks.

A particularly interesting modularity framework is Multislice modularity \cite{onnela}. The authors extend the null model of modularity (the random graph) to the novel multiplex (i.e. multidimensional) setting. They use several generalizations, namely an additional parameter that controls coupling between dimensions, basing their operation on the equivalence between modularity-like quality functions and Laplacian dynamics of populations of random walkers \cite{lambiotte-laplacian}. Basically they extend Lambiotte et al.'s work by allowing multidimensional paths for the random walker (\cite{signed-modularity}), considering the different connection types with different weights (\cite{barber-modularity}), and a different spread of these weights among the dimensions (\cite{traag-modularity}).

In order to represent both snapshots and dimensions of the network, the authors use slicing. Each slice $s$ of a network is represented by adjacency $A_{ijs}$ between nodes $i$ and $j$. The authors also specify inter-slice couplings $C_{jrs}$ that connect node $j$ in slice $r$ to itself in slice $s$. They notate the strengths of each node individually in each slice, so that  $k_{js} =  \sum_{i}  A_{ijs}$ and $c_{js} = \sum_{r} C_{jsr}$, and define the multislice strength $\kappa_{js} = k_{js} + c_{js}$. The authors then specify an associated multislice null model. The resulting multislice extended definition of modularity is the following: 
$$
Q = \dfrac{1}{2\mu}\sum_{ijsr}\left \{ \left ( A_{ijs} - \gamma_s \dfrac{k_{is}k_{js}}{2m_s}\delta_{sr}  \right ) + \delta_{ij}C_{jsr} \right \} \delta(c_{is}, c_{jr}).
$$
In this extension $\gamma_s$ is the resolution parameter, that may or may not be different for each slice. If $\gamma_s = 1$ for any $s$, then this formula degenerates on the usual interpretation of modularity as a count of the total weight of intra-slice edges minus the weight expected at random. Otherwise inter-slice coupling $C_{jsr}$ is considered. $C_{jsr}$ takes values from 0 to $\infty$. If $C_{jsr} = 0$ we degenerate again in the usual modularity definition. Otherwise the quality-optimizing partitions force the community assignment of a node to remain the same across all slices in which that node appears. In addition the multislice quality is reduced to that of an adjacency matrix summed over the contributions from the individual slices with a null model that respects the degree distributions of the individual contributions. The generality of this framework also enables different weights to be included across the $C_{jsr}$ couplings. After defining the new quality function, the algorithm needed to extract communities can be one of many modularity-based algorithms.

In Table 2 we merged all modularity approacches on the single ``Modularity'' row. One caveat is that, depending on the implementation, not all the features may be returned (for example only Multislice implementation is able to consider multidimensionality).

\subsection{MetaFac \cite{metafac}}
\begin{figure}
\centering
\includegraphics[scale=0.33]{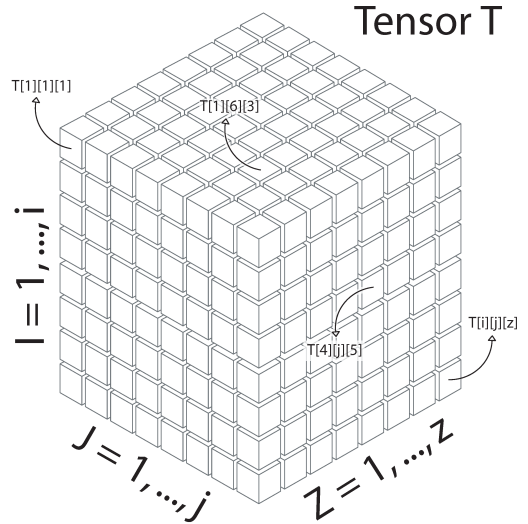}
\caption{A third-order tensor.} \label{fig:tensor}
\end{figure}

In this work, the authors introduce the concept of metagraph. The metagraph is a relational hypergraph to represent multi-relational and multi-dimensional social data. In practice, there are entities which connect to different kinds of objects in different ways (e.g.in a social media through tagging, commenting or publishing a photo, video or text). The aim is to discover a latent community structure in the metagraph, for example the common context of user actions in social media networks. In other words the authors are interested in clusters of people who interact with each other in a coherent manner. In this model, a set of entities of the same type is called a facet. An interaction between two or more facets is called a relation.

The idea of the authors is to use an $M$-way hyperedge to represent the interactions of $M$ facets: each facet as a vertex and each relation as a hyperedge on a hypergraph. A metagraph defines a particular structure of interactions between facets (groups of entities of the same type), not between facet elements (the entities themselves). In order to do so, the metagraph is defined as a set of data tensors. A tensor is an array with $N$ dimensions (see Figure \ref{fig:tensor} for an intuitive representation of a three dimensional tensor). This is a mathematical and computer science definition of tensors, for the notion of tensor in physics and engineering see \cite{tensor-real}. For an extensive review of tensors, tensor decomposition and their applications and tools see \cite{TensorReview} (in this work some examples are also provided of possible applications of tensor decompositions: signal processing \cite{tensor-signal}, numerical linear algebra \cite{tensor-algebra} and, closer to our area of interest, data mining \cite{tensor-dm1, tensor-dm2}, graph analysis tasks \cite{tensor-graph, tensorlinkprediction} and recommendation systems \cite{tensor-fact}).

Given the metagraph and its defined data tensors, the authors apply a tensor decomposition and factorization operation, which is a very hard task with a number of known issues. To the best of our knowledge, only recently have some memory and time efficient techniques been developed, such as \cite{tensor-decomposition}. In the metagraph approach the tensor decomposition can also be viewed as a dynamic analysis, when the sets of tensors are temporally annotated and the resulting core tensor refers to a specific time-step $t$. This is called metagraph factorization (for time evolving data).
Finally, the MF problem can be stated in terms of optimization, i.e. minimizing a given cost function, thus obtaining facet communities (for a time complexity of $\mathcal{O}(mnD)$).

\subsection{Variational Bayes \cite{hofman-bayesian}}
In this work, the authors model a complex network as a physical system, and then the problem of assigning each node to a module (inferring the hidden membership vector) in the network is tackled by solving the disorder-averaged partition function of a spin-glass.

The authors define a joint probability by considering the number of edges present and absent within and among the $K$ communities of a network. Traditional methods \cite{hastings06} need to specify $K$, this one is parameter free: the most probable number of modules (i.e. occupied spin states) is determined as $K^{∗} = argmax_{K} p(K|A)$. Such methods also need to infer posterior distributions over the model parameters (i.e. coupling constants and chemical potentials) $p(\pi, \theta|A)$ and the latent module assignments (i.e. spin states) $p(\sigma|A)$. The computationally intensive solution is tackled using the variational Bayes approach \cite{jordan-bv}.

This is a special case of the more general Stochastic Block Model, which is a family of solutions that reduces the community discovery problem to a statistical inference one. Historical approaches are \cite{historic-bayes-1, historic-bayes-2}, while other algorithms with the same technique, but different community definitions, are presented in different sections of this paper.

\subsection{$LA \rightarrow IS^{2}$* \cite{densityoverlap}}
In this work, the authors adopt the following definition of a community: a group $C$ of actors in a social network forms a community if its communication density function achieves a local maximum in the collection of groups that are close to $C$ \cite{densoverlap-base}. Basically, a group is a community if adding any new member to, or removing any current member from, the group decreases the average number of the communication exchanges.

This work is an evolution of \cite{old-densityoverlap}. It is built on two distinct phases: Link Aggregate (LA) and the real core of community detection ($IS^2$). The authors need a two-step approach because the $IS^2$ algorithm performs well at discovering communities given a good initial guess, for example when this guess is the output of another clustering algorithm, in this case called Link Aggregate (LA).

In LA, the nodes are ordered according to some criterion, for example decreasing Page Rank \cite{pagerank}, and then processed sequentially according to this ordering. A node is added to a cluster if adding it improves the cluster density. If the node is not added to any cluster, it creates a new cluster. The complexity of this stage is $\mathcal{O}(mk+n)$.

$IS^{2}$ explicitly constructs a cluster that is a local maximum w.r.t. a density metric by starting at a seed candidate cluster and updating it by adding or deleting one node at a time as long as the metric strictly improves. The  algorithm can be applied to the results of any other clustering technique, thus making this approach useful a general framework to improve some incomplete, or approximate, results.

\subsection{Local Density \cite{localdensity}}
In this work, the authors apply the classical approach which characterizes this category, i.e. to define a density quality measure to be optimized and then recursively merge clusters if this move produces an increase in the quality function. Here this function is the internal degree of a cluster $C$, i.e. the number of edges connecting vertices in $C$ to each other, $deg_{int}(\mathcal{C}) = |\{(u,v) \in E | u,v \in \mathcal{C}\}|$. Thus it is possible to define the local density of cluster as $$\delta_l(\mathcal{C})=\dfrac{2deg_{int}(\mathcal{C})}{|\mathcal{C}|(|\mathcal{C}|-1)}.$$
	
Optimizing $\delta \in [0, 1]$ alone makes small cliques superior to larger but slightly sparser sub-graphs, which is often impractical. For clusters to only have a few connections to the rest of the graph, one may optimize the relative density $$\delta_r(\mathcal{C}) = \dfrac{deg_{int}(\mathcal{C})}{deg_{int}(\mathcal{C}) + deg_{ext}(\mathcal{C})},$$ where $deg_{ext}(\mathcal{C}) = |\{(u,v) \in E | u \in \mathcal{C}, v \in V \setminus C\}|$. The final quality measure used is $f(\mathcal{C}) = \delta_l(\mathcal{C})\delta_r(\mathcal{C})$. A good approximation of the optimal cluster for a given vertex can be obtained by a local search, guided with simulated annealing \cite{simulated-annealing}.

\section{Bridge Detection}\label{sec:bridgecommunity}

\begin{figure}
\centering
\includegraphics[scale=0.6]{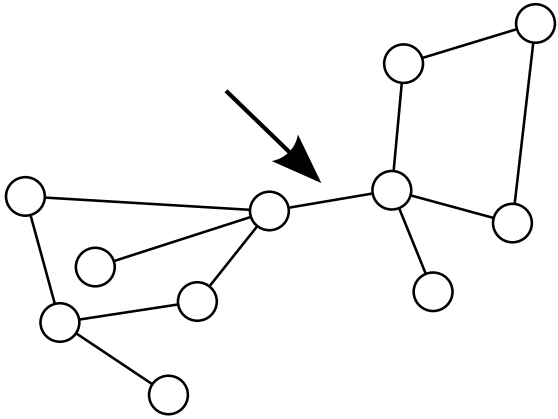}
\caption{An example of a graph that can be partitioned by identifying a ``bridge''.}\label{fig:metadef-bridge}
\end{figure}

The meta definition of community for the algorithms in this section is: 

\begin{definition}[Isolated Community]\label{def:bridgecommunity}
An isolated community in a complex network is a component of the network obtained by removing all the sparse bridges from the structure that connect the dense parts of the network.
\end{definition}

Usually, approaches in this category implement the following meta procedure:

\begin{procedure}
Rank nodes and edges in the network according to a measure of their contribution to keeping the network connected and then remove these bridges or avoid expanding the community by including them.
\end{procedure}

The bridge identified by the arrow in Figure \ref{fig:metadef-bridge} is a perfect example of an edge to be removed in order to decompose the network into disconnected components which represent our communities. The main focus for these approaches is how to find these bridges (which can be both nodes or edges) inside the network. The most popular approach in this category is to use a centrality measure. No assumptions at all are made about the internal density of the identified clusters.

In a social network analysis, a centrality measure is a metric defined in order to obtain a quantitative evaluation of the structural power of an entity in a network \cite{ucinet-book}. An entity does not have power in the abstract, it has power because it can dominate others. There are a number of measures defined to capture the power of an entity in a network. These include: Degree centrality, actors who have more ties to other actors may have more favorable positions; Closeness centrality, the closer an entity is to other entity in the network, the more power it has; Betweenness centrality, the most important entity in the network is the entity present in the majority of the shortest paths between all other entities.

Here we focus on two methods based on an edge definition of the traditional node betweenness centrality: the very first edge betweenness community discovery algorithm \cite{edgebetween}, which has recently been the focus of further evolutions, i.e. a general approach that uses split betweenness in order to obtain an overlapping community discovery framework \cite{conga2}. We then also consider two alternative methods \cite{lshell, lancichinetti} which try to detect the bridges by expanding the community structure and computing a community fitness function.

As can be seen in Table 2, these algorithms are good at finding overlapping partitions (it is not the original edge betweenness algorithm, however basically the CONGA strategy enables it to detect overlapping clusters). The weak points of this approach appear when dealing with dynamic, multidimensional or incremental structures. We are not able to prove this point in the experimental section so we will use an intuitive explanation. In order to compute the fitness function to detect bridges, it is necessary to start from the assumption that the algorithm is a complete representation of all connections among the clusters, which may be hard in an incremental setting. Furthermore, for routing algorithms that are needed to compute the betweenness or closeness centrality, there are some constraints on the structure of the network which are not satisfied in a multidimensional setting. Consider a network with two dimensions and a rule that states that jumping from one dimension to another, lowers the cost of the path. We thus have negative cycles and a significant shortest path cannot be computed (since in Bellman-Ford's algorithm, disallowing edge repetition, it is possible to obtain a shortest path that will always cross all the negative cycles it can, thus destroying the concept of bridge \cite{bellmanford}).

\begin{figure}
\centering
\includegraphics[scale=0.4]{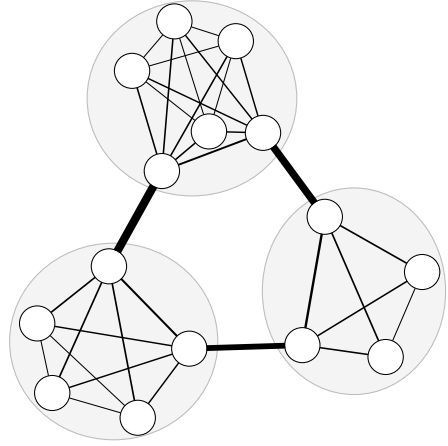}
\caption{An intuitive example of the bridge detection approach. In this graph the edge width is proportional to the edge betweenness value. Wider edges are more likely to be a bridge between communities.} \label{fig:edgebet}
\end{figure}

\subsection{Edge Betweenness \cite{edgebetween}}\label{sec:edgebet}
The main assumption of this work is that if a network contains communities or groups that are only loosely connected by a few inter-group edges, then all the shortest paths between different communities must go along one of these edges. In order to find these edges, which are mostly between other pairs of vertices, the authors generalize Freeman's betweenness centrality \cite{betweennesscentr} to edges, and define the ``edge betweenness'' of an edge as the number of shortest paths between pairs of vertices that run along it. Figure \ref{fig:edgebet} depicts an example, where the size of the edges is proportional to their edge betweenness. As can be seen, the higher edge betweenness values are taken by the edges between communities. By removing these edges, it is possible to separate one group from one another and thus reveal the underlying community structure of the graph.

This is one of the first community discovery algorithms developed after the renewed interest in social network analysis that started in the late 1990s. Previously the traditional graph partitioning approach constructed communities by adding the strongest edges to an initially empty vertex set (as in hierarchical clustering \cite{hierarchicalclustering}). Here, the authors construct communities by progressively removing edges from the original graph.

While the classical implementation of the edge betweenness algorithm is $\mathcal{O}(mn)$, a speed-up for parallel systems that are linear [106] has recently been proposed. Thus without the parallel algorithm the worst case time complexity is $\mathcal{O}(m^2n)$. There are slight variations of this method using different centrality measures (\cite{edgeclustering, loopcoefficient}).

\subsection{CONGA \cite{conga2}}
CONGA (Cluster-Overlap Newman Girvan Algorithm) is based on the well-known edge betweenness community discovery algorithm \cite{edgebetween}, described in Section \ref{sec:edgebet}. It adds the ability to split vertices between communities, based on the new concept of ``split betweenness''.

The split betweenness \cite{splitbetween} of a vertex $v$ is the number of shortest paths that would pass between the two parts of $v$ if it was split. There are many ways to split a vertex into two, the best split is the one that maximizes the split betweenness. Basically, with the following split operation, any disjoint community discovery algorithm can be applied and returns overlapping partitions (\cite{conga}):

\begin{enumerate}
\item Calculate edge betweenness of edges and split betweenness of vertices. 
\item Remove edge with maximum edge betweenness or split vertex with maximum split betweenness, if greater. 
\item Recalculate edge betweenness and split betweenness. 
\item Repeat from step 2 until no edges remain.
\end{enumerate} 

Given a relaxed assumption on the edge betweenness computation, the total time complexity of CONGA is $\mathcal{O}(n \log n)$.

\subsection{L-Shell \cite{lshell}}
In L-Shell algorithm, the idea is to expand a community as much as it can, stopping the expansion whenever the network structure does not allow any further expansion, i.e. the bridges are reached.

The key concept is the $l-shell$, a group of $l$ vertices whose aim is to grow and occupy an entire community while two quantities are computed: the emerging degree and total emerging degree. The emerging degree of a vertex is defined as the number of edges that connect that vertex to vertices that the $l-shell$ has not already visited as it expanded from the previous $(l-1)$, $(l-2)$, ... $-shells$. The total emerging degree $K_{j}$ of an $l-shell$ is thus the sum of the emerging degrees of all vertices on the leading edge of the $l−shell$.

For a starting vertex $j$ the algorithm starts an $l-shell$, $l=0$, at vertex $j$ (add $j$ to the list of community members) and computes the total emerging degree of the shell. Then it spreads the $l-shell$, $l=1$, it adds the neighbors of $j$ to the list, and computes the new total emerging degree. Now it can compute the change in the emerging degree of the shell. If the total emerging degree is increased less than a given threshold $\alpha$, then a community has been found. Otherwise it increases the size of the shell (posing $l=l+1$) until $\alpha$ is crossed or the entire connected component is added to the community list. As can be seen, for each node we have a quadratic problem, i.e. the time complexity is $\mathcal{O}(n^3)$. The assumption is that a community is a structure in which the total emerging degree cannot be significantly increased, i.e. the vertices at the border of the community have few edges outside it and these edges are the bridges among different communities.

\subsection{Internal-External Degree \cite{lancichinetti}}
An approach close to $l-shell$ starts from the similar basic assumption that communities are essentially local structures, involving the nodes belonging to the modules themselves plus at most an extended neighborhood of them. The fitness chosen here is the total internal degree of nodes on the sum of internal and external degrees to the power of a positive real-valued parameter ($\alpha$). Given a fitness function, the fitness of a node $A$ with respect to sub-graph $\mathcal{G}$, $f_{G}$, is defined as the variation of the fitness of sub-graph $\mathcal{G}$ with and without node $A$. The process of calculating the fitness of the nodes and them joining them together in a community stops when the nodes examined in the neighborhood of $\mathcal{G}$ all have negative fitness, i.e. their external edges are all bridges, after a total time complexity of $\mathcal{O}(n^2 \log n)$.

Large values of $\alpha$ yield very small communities, instead small values deliver large modules. For $\alpha$=1 this method recalls \cite{edgeclustering} closely, which is another algorithm that falls into this category. Going from $\alpha$=0.5 to $\alpha$=2 reveals the hierarchical structure of the network.

\section{Diffusion}\label{sec:percolation}

\begin{figure}
\centering
\includegraphics[scale=0.6]{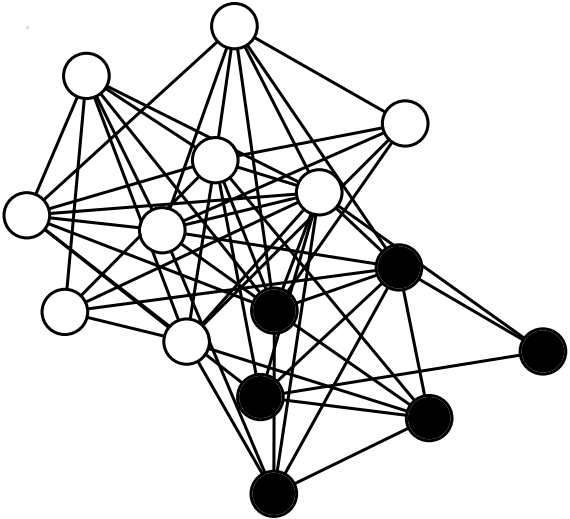}
\caption{An example of graph partitioned with a diffusion process.}\label{fig:metadef-diffusion}
\end{figure}

A diffusion is a process in which vertices or edges of a graph are randomly designated as either ``occupied'' or ``unoccupied'' the various properties of the resulting patterns of vertices are then queried \cite{newman-structurefunction} (see Figure \ref{fig:metadef-diffusion}, which also highlights the lack of clear bridges between communities or any density difference between the inside and the outside of clusters). A generalization of a diffusion process can be used for community discovery in complex networks, according to the following definition of community: 

\begin{definition}[Diffusion Community]\label{def:diffcommunity}
A diffusion community in a complex network is a set of nodes that are grouped together by the propagation of the same property, action or information in the network. 
\end{definition}

The definition of the meta procedure followed by algorithms in this category is thus:

\begin{procedure}
Perform a diffusion or percolation procedure on the network following a particular set of transmission rules and then group together any nodes that end up in the same state.
\end{procedure}

According to this meta definition, a community can also be defined as a set of entities influenced by a fixed set of sources. This is important because algorithms, which are not explicitly developed as approaches for graph partitioning, are also considered as a community discovery method. Basically, this definition of the problem overlaps with another well-known data mining problem: influence spread and flow maximization \cite{flake-flow}, which is often used for viral marketing \cite{viralmarketing}. Preliminary ideas can be found in \cite{fortunato-infocen}, even if only a novel centrality measure is defined, and the approach can be mapped in the Newman edge betweenness algorithm \cite{edgebetween}. Another approach that mixes physics and information theory is \cite{infotheo-mod}.

Other interesting works in viral marketing are, given a community partition, the analysis of the group characteristics in order to predict their evolution \cite{burst}. In addition, it is possible to predict if a single vertex will be attached to a group, or even classify some features (and the evolution of these features) of a group. While it is not a community discovery work, \cite{burst} can be used as a framework after a community detection algorithm in order to obtain a temporal evolving description of the identified groups.

To sum up, the classical community discovery diffusion-based algorithms presented here are: a label propagation technique \cite{labelprop}, dynamic node coloring for temporal evolving communities \cite{node-coloring}, and edge resistor algorithms that consider the original graph as an electric circuit \cite{kirchhoff}.

The influence propagation approaches reviewed here are: an analytical description of a network representing an exchange of information \cite{commdyn}; GuruMine \cite{gurumine}, a framework whose aim is to analyze ``tribes'', DegreeDiscountIC \cite{flu-maxim}, a classical spread maximization algorithm, and a mixed membership stochastic blockmodel algorithm \cite{mixedmembership}, which uses Bayesian inferences in order to compute the final state of the influence vectors for each node in the network.

In this category, it is natural to deal with directed communities, since the diffusion process, when dealing with information spread, is naturally modeled following asymmetric relations. It is also intrinsically dynamic, thus many diffusion algorithms provide this feature in the community discovery solution. We found that no approach currently considers multidimensional networks, however we believe that considering different communication channels inside a network should be a key feature of this category.

\begin{figure}
\centering
\includegraphics[scale=0.19]{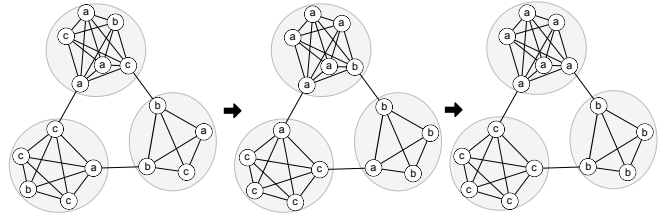}
\caption{Possible steps of a label propagation-based community discoverer.} \label{fig:labelprop}
\end{figure}

\subsection{Label Propagation \cite{labelprop}}
Suppose that a node $x$ has neighbors $x_{1}, x_{2}, ... , x_{k}$ and that each neighbor carries a label denoting the community that it belongs to. Then $x$ determines its community based on the labels of its neighbors. A three-step example of this principle is shown in Figure \ref{fig:labelprop}.

The authors assume that each node in the network chooses to join the community to which the maximum number of its neighbors belong. As the labels propagate, densely connected groups of nodes quickly reach a consensus on a unique label. At the end of the propagation process, after a quasi-linear time complexity ($\mathcal{O}(m+n)$) nodes with the same labels are grouped together as one community.

Clearly, a node with an equal maximum number of neighbors in two or more communities can belong to both communities, thus identifying overlapping communities. It is easy to define an overlapping version of this algorithm \cite{labeloverlapping}.

\subsection{Node coloring \cite{node-coloring}}
Consider an affiliation network in which some individuals form groups by attending the same event. In this approach, which represents an evolution of \cite{prev-node-color}, the base input representation is an evolving bipartite graph of individuals connected to events.

Various rules have been defined to connect groups over time and form communities of groups: 
\begin{enumerate}
\item In each time step, every group is a representative of a distinct community; 
\item An individual is a member of exactly one community at any one time (but can change community affiliation over time); 
\item An individual tends not to change his / her community affiliation very frequently; 
\item If an individual keeps changing affiliations from one community to another, then it is not a true member of any of those communities; 
\item An individual is frequently present in the group representing the community with which he / she is affiliated. 
\end{enumerate}

The authors define the community interpretation of a graph $G$ as a function $f : V \rightarrow \mathbb{N}$. Each individual belongs to exactly one community in each time-step, and each group represents exactly one community. Thus, although the affiliation can change over time, this is a disjoint community detection algorithm, not an overlapping one.
To measure the quality of a community interpretation, the authors use costs (whenever an individual changes color, or it connects to groups with different colors, and so on) to penalize violations of Rules 3 and 5. The optimization problem is then to find the valid community interpretation by minimizing the total cost resulting from the individual edges, group edges and color usage.
The authors present an exhaustive global optimum algorithm with exponential time complexity (the algorithm with dynamic programming tries all possible colorings of the graph) and then some heuristics, ending up with a final complexity of $\mathcal{O}(ntk^2)$. In \cite{node-coloring2} the authors present another set of heuristics and optimizations.

\subsection{Kirchhoff \cite{kirchhoff}}
In this paper, the basic idea is to imagine each edge as a resistor with the same resistance. It is then possible to connect a virtual ``battery'' between chosen vertices so that they have fixed voltages. Having made these assumptions the graph can be viewed as an electric circuit with a current flowing through each edge (resistor). By solving Kirchhoff's equations, the authors obtain the voltage value of each node. The authors claim that, from a node's voltage value they are able to judge whether it belongs to one community or another. This approach is very efficient, since the complexity is $\mathcal{O}(m+n)$.

A further expansion \cite{kirch2} applies a walk-based approach in order to unveil the hidden hierarchical structure of the network and identify good choices for the seed poles. The authors then apply a very similar implementation of this method using a Kirchhoff matrix.

\subsection{Communication Dynamic \cite{commdyn}}
The focus of the paper is an analytical description of the evolution of a network, whose size is stable over time and represents the exchange of communication among individuals. The authors present a locality based model for communication dynamics, which can be used in order to identify the mechanisms of community creation and evolution over time in a social network.

Similar approaches, such as preferential attachment \cite{barabasi-pa}, are applicable only when communications are open (observable to all nodes). Instead the authors present a locality based model which relies on two fundamental principles: firstly the concept of locality reduces the set of nodes that a node can attach to in the next time step. Secondly, after obtaining a node's locality, the attachment mechanism, which is used by the individual to select the nodes in its locality to which it will connect at the next time step, must be specified. This is a Markov chain-like approach.

For the preliminary community structure that identifies the local environment of a node, the authors use an existing method based on density \cite{densityoverlap}. The authors define the blogograph as a directed, unweighted graph representing the communication of the blog network within a fixed time-period. There is a vertex in the blogograph representing each blogger and a directed edge from the author of any comment to the owner of the blog. The authors consider consecutive weekly snapshots of the network.

The authors recorded statistics as numbers of vertices, edges, the power-law degree distribution exponent, giant component size and so on, observing that they are stable over time, consistent with previous observations (as in \cite{barabasi-evo} and \cite{watts-stats}). They also provide an indicator of community vitality over time. The goal is to produce a sequence of graphs which simulate the connection and reconnection of vertices and can be used for community validation.

\subsection{GuruMine \cite{gurumine}}

\begin{figure}
\centering
\subfigure[Action Table]{
\scalebox{0.7}{
\begin{tabular}[b]{|ccc|}
 \hline
User & Action & Time \\
\hline
U1 & $\beta$ & 12 \\
U5 & $\beta$ & 14 \\
U1 & $\alpha$ & 15 \\
U2 & $\beta$ & 15 \\
U3 & $\beta$ & 16 \\
U4 & $\beta$ & 17 \\
U2 & $\alpha$ & 18 \\
U4 & $\alpha$ & 19 \\
U3 & $\alpha$ & 19 \\
\hline
\end{tabular}\label{fig:gurutable}
}
}
\subfigure[Action $\alpha$]{\includegraphics[scale=0.25]{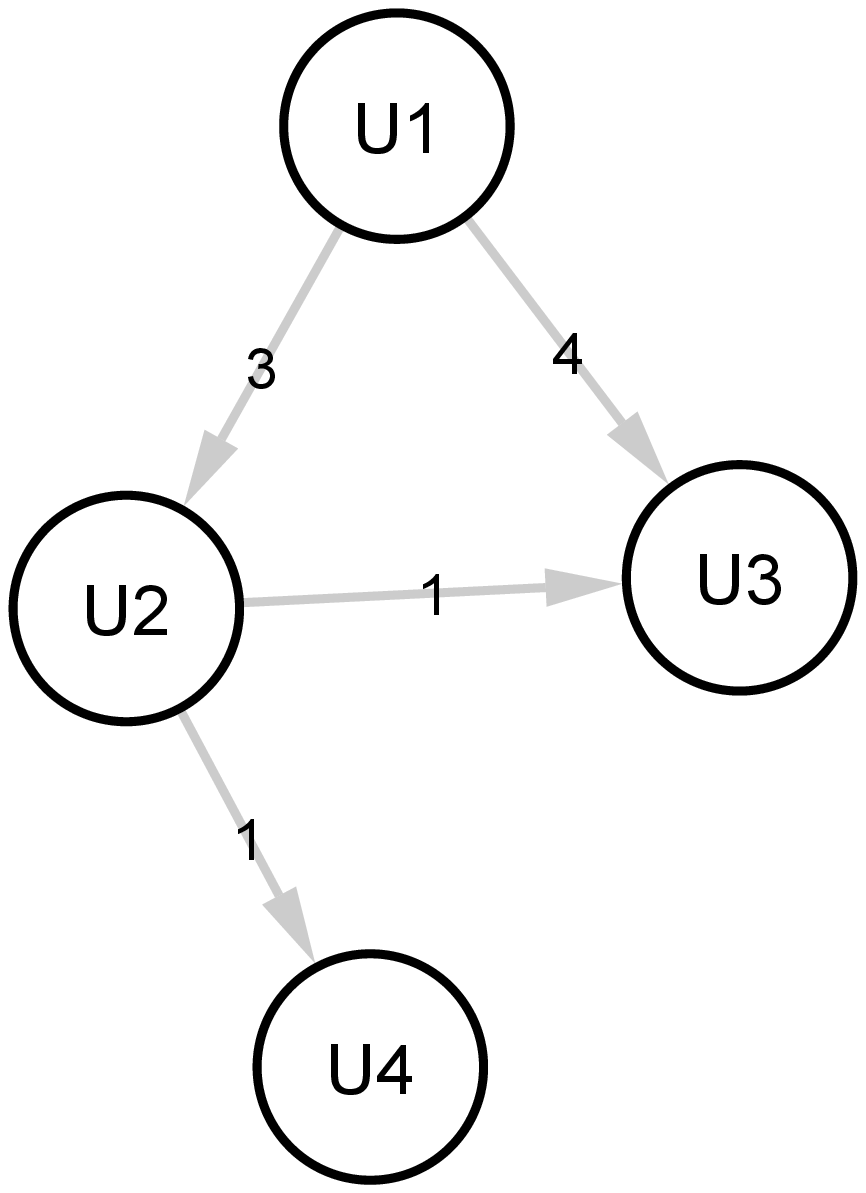}\label{fig:guru-actionalpha}}
\subfigure[Action $\beta$]{\includegraphics[scale=0.25]{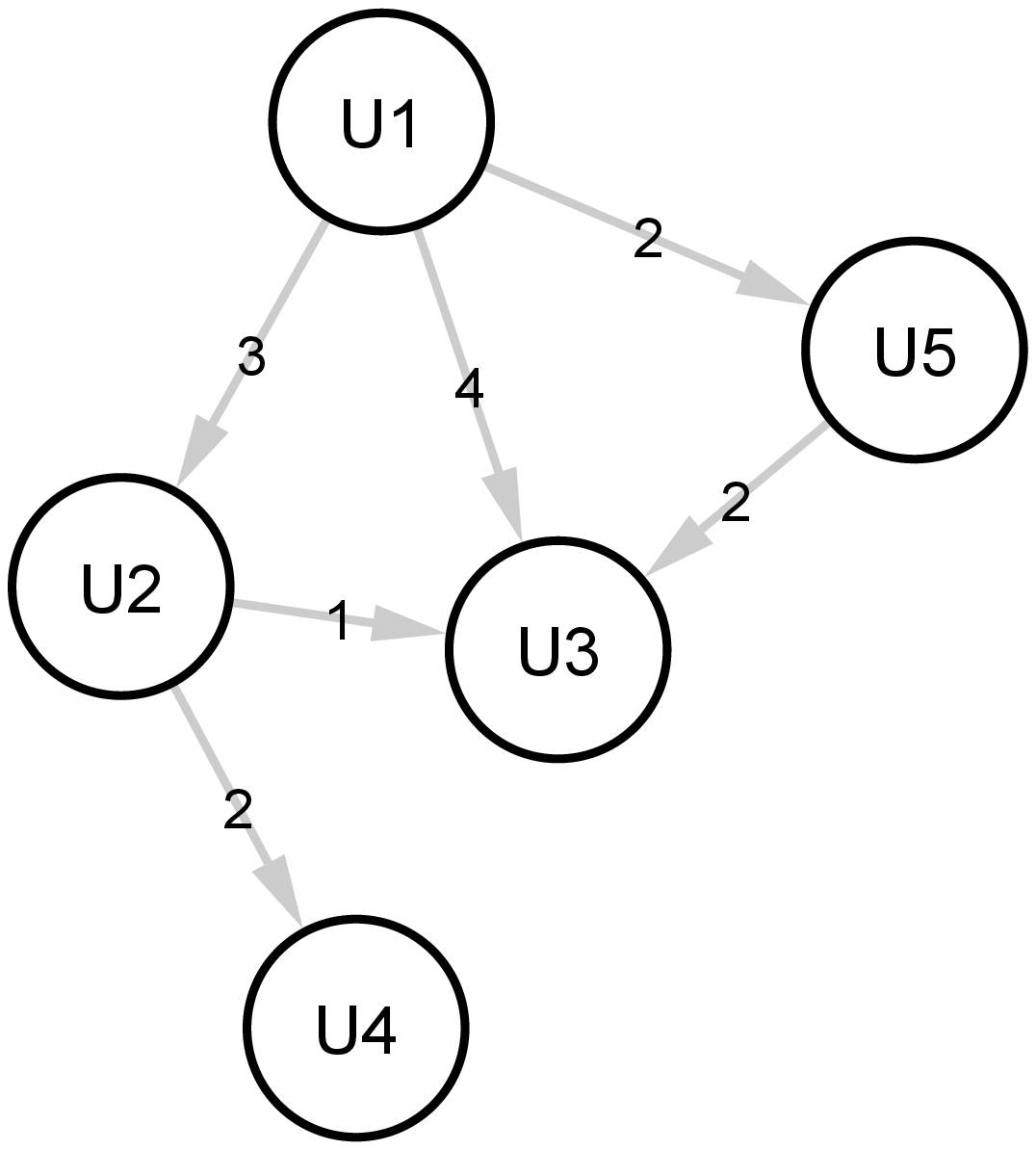}\label{fig:guru-actionbeta}}
\caption{The GuruMine data structures: the action table and the influence graphs.} 
\end{figure}

The aim of GuruMine is to investigate how influence (for performing certain actions) propagates from users to their network friends, potentially recursively, thus identifying a group of users that behave homogeneously (i.e. a tribe, or a community). For instance, Table 3 shows a possible action table with two actions, $\alpha$ and $\beta$, and five users. Figures \ref{fig:guru-actionalpha} and \ref{fig:guru-actionbeta} represent the influence graphs of these two actions. $U1$ can be considered as a tribe leader in both cases. However, for action $\alpha$, $U1$ cannot be considered a leader if the threshold regarding the minimum number of influenced users is equal to 4.

Since the set of influenced users is the same, we have a ``tribe leader'', meaning the user leads a fixed set of users (tribe) w.r.t. a set of actions, which can be considered a community. The general goal is similar to recent works such as \cite{influblog, leskoveccost, kempemax}. However, here the input includes not just a graph (which is not edge-weighted) but also an action table which plays a central role in the definition of leaders. This action table contains a triple $(u; t; a)$ indicating that user $u$ performed action $a$ at time $t$, from which a directed propagation graph is derived. If the composition of the influenced graph is the same, we have a tribe.

Any algorithm for extracting leaders must scan the action log table and traverse the graph (which means that the complexity also depends on this table and is $\mathcal{O}(TAn^2)$). The implementation works with only one scan, with the action log stored in chronological order. With this scan the influence matrix $IM_{\pi}(U; A)$ can be computed. For tribe leaders the influence cube $Users \times Actions \times Users$ is needed, with cells containing Boolean entries if user $v$ was influenced by user $u$ w.r.t. action $a$. A tribe is essentially an item-set, i.e. a community with common behavior. This phase is implemented by ExAMiner \cite{examiner}. This work is part of a larger framework that also has a query interface \cite{gurumine2}.

\subsection{DegreeDiscountIC \cite{flu-maxim}}
This work is in the context of the classical data mining influence spread. The problem definition consists in deciding who to include in the initial set of targeted users so that, if necessary, they influence the largest number of people in the network. This knowledge can be used for community discovery: each seed node is the head of a community that acts uniformly, and the set of these influenced nodes is the community members. This work is an implementation of the idea in \cite{kempemax} and the improvement of the algorithm proposed in \cite{leskoveccost}.

Influence is propagated in the classical network representation of social interactions according to a stochastic cascade model. Let S be the subset of vertices selected to initiate the influence propagation. In the cascade model (IC), let $A_i$ be the set of vertices that are activated in the $i$-th round, and $A_0 = S$. For each edge with one inactive endpoint, there is a probability of activation proportional to the active neighbors, and this is repeated until the cascade cannot expand any further. Then all edges not used for propagation are removed, and the set of influenced vertices is simply the set of vertices reachable from $S$ in $G'$. This cascade can be evolved in a weighted model (WC), by considering the number of inactive neighbors of an active node and the activated neighbors of an inactive node. A discount on the degree of these vertices is considered if both connected nodes are part of the seed set. With this and more finely tuned heuristics on degrees, the authors manage to develop a well performing algorithm with a reasonable level of complexity (equal to $\mathcal{O}(k\log n + m)$).

\subsection{MMSB \cite{mixedmembership}}
In the mixed membership stochastic blockmodel approach (MMSB), the authors implement the following mechanism: each node belongs to any possible community with a certain probability. These probabilities are then influenced by the probabilities of all other nodes. In practice, the influence of affiliations spreads over the network until convergence, by averaging the vector of probabilities of each node with the vector of the general influences. In other words, this process is equivalent to label propagation, and instead of a simple number indicating the membership there is a vector of probabilities.

The indicator vectors are in the form of $\overrightarrow{z}_{p \rightarrow q}$, which denotes the group membership of node $p$ when it is approached by node $q$ (note that this is not symmetric). Then, for each node $i$ a mixed membership vector $\overrightarrow{\pi_i}$ is drawn, and the value of the interaction between this vector and the original one of the node is sampled. The authors also introduce a sparsity parameter to calibrate the importance of non-interaction.

As for other mixed membership models, this is intractable to compute. A number of approximate inference algorithms for mixed membership models have recently appeared such as mean-field variational methods \cite{bayesian1}, expectation propagation \cite{expprop} and Monte Carlo Markov chain sampling \cite{bayesian2}. In these papers, the authors apply mean-field variational methods to approximate the posterior of interest, which has a complexity of $\mathcal{O}(nk)$. An extension of this work which considers also the degree of the vertices as a normalization factor is \cite{MMSB-degree}. A work very related to this one, working with a very similar notion of propagating probabilities as influence or information, is \cite{MMSB-2}.

\section{Closeness}\label{sec:closeness}

\begin{figure}
\centering
\includegraphics[scale=0.6]{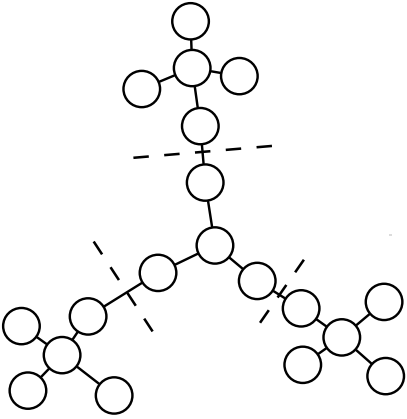}
\caption{An example of a graph which can be partitioned by considering the relative distance, in terms of number of edges, among its vertices.}\label{fig:metadef-closeness}
\end{figure}

A very intuitive notion of community in a complex network is based on the concept of how close its members are connected together. A community is a set of individuals who can communicate with each other very easily because they can reach any other member in a relatively lower number of hops than the network's average. Figure \ref{fig:metadef-closeness} shows a simple example of this configuration. The underlying definition of community in this case is:

\begin{definition}[Small World Community]\label{def:closecommunity}
A small world community in a complex network is a set of nodes that can reach any member of its group usually by crossing a very low number of edges, significantly lower than the average shortest path in the network.
\end{definition}

We use the term ``small world'' \cite{watts-smallworld} since it conveys the idea of very closely connected nodes. A very efficient approach used with this problem definition relies on random walks. A random walk is a process in which at each time step a walker is on a vertex and moves to a vertex chosen randomly and uniformly from its neighbors. The same procedure is followed for the new selected vertex. This is a Markov process. However, various strategies have been formulated in order to obtain very sophisticated random walk based application. For example, the popular link analysis PageRank algorithm \cite{pagerank} is based on random walks. This ends up in the following meta procedure:

\begin{procedure}
Given a network, perform several random walks and then cluster together nodes which appear frequently in the same walk. 
\end{procedure}

Algorithms in this category inherit the weakness in multidimensional networks from Bridge Detection algorithms, since also in this case paths are important in this community discovery category.

To the best of our knowledge there are three main community discoverers that use random walks in order to find communities whose members are very close to each other: Walktrap \cite{walktrap}, based on the assumption that when performing random walks the virtual surfer is trapped in the high density regions of the graph (i.e. the communities); DOCS \cite{docs}, a more complex framework that also uses modularity as a fitness function; and Infomap \cite{infomap}, which applies an information-theoretic approach. An older approach in this category is the Markov Cluster Algorithm \cite{vanDongen2000}, which is still commonly used especially in bioinformatics. It simulates a controlled flow through random walks in a network using matrix multiplication and inflation.

\subsection{Walktrap \cite{walktrap}}
The Walktrap approach is based on the following intuition: random walks are able to unveil the real distance among nodes by frequently exploring nodes in the same community. The key problem is the definition of the distance function between any two vertices, computed from the information given by random walks in the graph. High values of this measure mean that the two vertices $i$ and $j$ ``see'' the network in a very similar way, thus they belong to the same community. Therefore, this distance must be large if the two vertices are in different communities, and small otherwise. In the original paper this distance is defined as: 
$$r_{ij} = \sqrt{\sum_{k = 1}^{n} \frac{(P_{ik}^{t} - P_{jk}^{t})^2}{d(k)}}$$
where $P_{ik}^{t}$ is the probability to go from $i$ to $j$ in $t$ steps and $d(k)$ is the degree of vertex $k$.

A critical parameter is the length $t$ of the random walks: it must be sufficiently long to gather enough information regarding the topology of the graph. However it must not be too long because when the length of a random walk starting at vertex $i$ tends towards infinity, the probability of being on a vertex $j$ only depends on the degree of vertex $j$ (and not on the starting vertex i).

Similar random walk approaches are \cite{randomw-1, randomw-2}. However they are less efficient compared to the average complexity of Walktrap, which is at the worst case $\mathcal{O}(mn^2)$.

\subsection{DOCS \cite{docs}}
This method is based on a spectral partition and random walk expansion, and is an extension of \cite{prevdocs}. The general idea is to obtain an initial guess in a first step regarding the community structure, and then collapse or expand these communities according to the hints given by the random walks among them. 

The first step is to coarsen the original graph into a series of higher level graphs. This is guided by modularity maximization. In the lazy random walk stage, vertices are labeled as contributing or non contributing vertices depending on whether or not they can be moved to another cluster and provide an increase in modularity. They are also sorted in a descending order by their contributing values. The target communities can then be extracted.

\subsection{Infomap \cite{infomap}}
The Infomap algorithm is one of the most accurate community discovery methods \cite{lanci-comparative}. It is based on a combination of information-theoretic techniques and random walks. The authors explore the graph structure with a number of random walks of a given length and with a given probability of jumping to a random node. This approach is equivalent to the random surfer of the PageRank algorithm \cite{pagerank}.

Intuitively, the random walkers are trapped in a community and exit from it very rarely. Each walk is described as a sequence of steps inside a community followed by a jump. By using unique names for communities and reusing a short code for nodes inside the community, this description can be highly compressed, in the same way as re-using street names (nodes) inside different cities (communities). The renaming is done by assigning a Huffman coding to the nodes of the network. The best network partition will result in the shortest description for all the walks.

\section{Structure Definition}\label{sec:structure}
A number of works tackle community discovery with a very strong assumption: to be called a community, a group of vertices must follow a very strict structural property. In other words, they use the following meta definition of community:

\begin{definition}[Structure Community]\label{def:structuralcommunity}
A structure community in a complex network is a set of nodes with a precise number of edges between them, distributed in a very precise topology defined by a number of rules. Sets of nodes that do not satisfy these structural rules are not communities. 
\end{definition}

The aim of the community discovery algorithm is to find all the maximal structures in the network that satisfy the desired constraints. The corresponding meta procedure implemented in this category is simple (i.e. find in a efficient way all the maximal structure defined) and hence there is no need to discuss it further.

This task is similar to a very well-known data mining problem in network analysis: graph mining. Some examples of graph mining algorithms are \cite{gspan, germ, gaston, singlegraphminer}. However, traditional graph mining algorithms only return all the single different structure patterns with their support. In community discovery there is only one important structure and the desired result is the list of all vertex groups that make up that structure in the network.

We will thus ignore pure graph mining algorithms and just focus on structural community discovery approaches. The methods reviewed here are: clique percolation \cite{kclique} and its evolution for bipartite graphs \cite{biclique}, the s-plexes detection \cite{splexes} and a maximal clique approach \cite{eagle}. We will not focus on other minor evolutions, such as the k-dense approaches \cite{kdense-jap}.

Since a defined structure may be, without any constraint, overlapping, weighted, directed or multidimensional, there is virtually no structural feature that cannot be embedded in a definition used by the algorithms in this category. Depending on the desired structure, analysts can also find communities that do not overlap with any of the previous categories, thus avoiding densities, or bridges or any other previous definition. The downside of this strategy arises when working in an incremental setting: given a simple modification on the structure, such as adding or deleting a single node or edge, the algorithm is likely to recompute everything from scratch. This is because properties of the substructure that are discovered may be violated by any single modification.

\subsection{K-Cliques \cite{kclique}}

\begin{figure}
\centering
\includegraphics[scale=0.3]{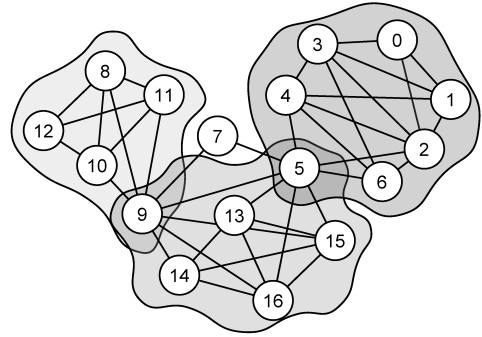}
\caption{The overlapping community structure detected by a clique-percolation approach.} \label{fig:cliqueperc}
\end{figure}

Palla et al. suggest that a community can be interpreted as a union of smaller complete (fully connected) sub-graphs that share nodes. The authors define a k-clique-community as the union of all k-cliques that can be reached from each other through a series of adjacent k-cliques. Two k-cliques are said to be adjacent if they share $k - 1$ nodes. A 2-clique is simply an edge and a 2-clique-community is the union of those edges that can be reached from each other through a series of shared nodes. Consider Figure \ref{fig:cliqueperc}. In this case the clique percolation approach detects $\{0,1,2,3\}$ as a 4-clique. Then it considers $\{1,2,3,4\}$: it is again a 4-clique and it shares 3 vertices with the previous one. Thus the two cliques are joined in one community. The same is true for the 4-cliques $\{2,3,4,6\}$ and $\{2,4,5,6\}$, thus identifying the community $\{0,1,2,3,4,5,6\}$. In this process, two communities can have an overlap of some vertices (in the example, vertices 5 and 9).

The algorithm first extracts all complete sub-graphs of the network that are not part of a larger complete sub-graph. The aim of the first phase is to populate a clique-clique overlap matrix. In this data structure each row (and column) represents a clique and the matrix elements are equal to the number of common nodes between the corresponding two cliques. The diagonal entries are equal to the size of the clique. The k-clique-communities can be found by erasing every off-diagonal entry smaller than $k - 1$. The complexity of this procedure, since the hardness of clique detection, is $\mathcal{O}(m^{\frac{\ln m}{10}})$.

\subsection{S-Plexes Enumeration \cite{splexes}}
An s-plex is a relaxed concept of the c-isolated clique \cite{cisolated, cisolated2}. Let $G = (V, E)$ be an undirected graph. A set $S \subseteq V$ of $k$ vertices is called c-isolated if it has less than $ck$ outgoing edges, where an outgoing edge is an edge between a vertex in $S$ and a vertex in $V \backslash S$. A c-isolated clique is a concept that is considered too restrictive for a community. Instead, the authors use a relaxed version of a c-isolated clique called s-plex \cite{splextheory}: in an undirected graph $G = (V, E)$, a vertex subset $S \subseteq V$ of size $k$ is called an s-plex if the minimum degree in $G[S]$ is at least $k - s$. Hence, cliques are exactly 1-plexes.

Since in an s-plex $S$ of size $k$ every vertex $v \in S$ is adjacent to at least $k - s$ vertices, the sub-graph induced by $S$ in the complement graph (the graph with the same set of vertices and complementary edge set) $G[S]$ is a graph with a maximum degree of at most $s - 1$. The idea is to enumerate maximal s-plexes in $G$ by deleting minimal sub-graphs with a maximal degree of $s - 1$ in the complement graph. A key concept for this solution is the pivot set $P$. The pivot set contains the pivot vertex $v$ and those vertices that belong to the s-plex but are not adjacent to $v$. The pivot vertex is defined as the vertex with the lowest index of those vertices with less than $c$ outgoing edges.

The algorithm is an evolution of \cite{oldsplex} and removes vertices from the candidate set $C$ with too few neighbors in $C$. It builds the complement graph, then for each possible pivot set $P$ applies the deletion of minimal sub-graph in the complement graph. Finally, it removes enumerated s-plexes that either have pivot $u \neq v$ or are not maximal. The complexity is $\mathcal{O}(knm)$.

\subsection{Bi-Clique \cite{biclique}}
This is a bipartite graph version that solves various issues regarding the k-clique approach \cite{kclique}, namely the impossibility to analyze sparse network regions, due to the fact that 2-clique communities are simply the connected components of the network. The first non-trivial k-clique has size k = 3 and nodes must have at least two links in order to qualify for participation in a 3-clique. In networks with heavy tailed degree distributions, a large fraction of the nodes have less than two edges.

Bi-clique is a natural approach for affiliation networks, where in a one-mode projection all (sparse) information regarding the bipartite linkages is reduced to a giant quasi-clique. All the information contained in edge weights is typically discarded in a subsequent thresholding operation. The Bi-Clique algorithm detects structures between 2-clique communities and 3-clique communities where the k-clique algorithm usually fails.

The algorithm begins by isolating the $N$ maximal bi-cliques in the bipartite network using \cite{lcm3}. Using this list the authors create two symmetric clique overlap matrixes for the two classes of nodes. Then, for both matrix diagonal, elements greater than or equal to $a$ and $b$ (the two parameters of the algorithm) respectively are set to one, while everything else is set to zero. The final overlapping matrix is obtained by the matrix intersection, using the AND operator. The final step is to determine the connected components of $L$; each component corresponds to a bi-clique community. The final complexity of the approach is $\mathcal{O}(m^2)$.

\subsection{EAGLE \cite{eagle}}
EAGLE starts from the following assumption: in every dense-linked community there is at least one large clique. This clique could be considered the core of the community. EAGLE firstly finds out all the maximal cliques in the network with the Bron-Kerbosch algorithm \cite{cliqdetect} (complexity $\mathcal{O}(3^{\frac{n}{3}})$), discarding those whose vertices are part of other larger maximal cliques and those with less than $k$ vertices. EAGLE then calculates the similarity between each pair of communities. It then selects the pair of communities with the maximum similarity, incorporating them into a new community and calculating the similarity between the new community and other communities. The similarity measure is the modularity \cite{clauset-modularity}. This calculation is repeated until only one community remains, thus completing the dendrogram. 

The second stage is to cut the dendrogram. Any cut through the dendrogram produces a cover of the network. To determine the place of the cut, a measurement is required to judge the quality of a cover, computed with a given variant of modularity.

\section{Link Clustering}\label{sec:linkcommunity}
Some recent approaches have been based on the idea that the community is not a partition of network nodes, but a partition of the links. In other words, it is the relationship between two entities that belongs to a particular environment and the entities belong to all the communities of their edges (or a subset of them).

The meta procedure in this class is:

\begin{procedure}
We are given a set of relations M between a set of entities N. We cluster together relations that are similar, i.e. established between the same set of entities, and we then connect each entity n to the communities its relations belong to. 
\end{procedure}

The underlying meta definition of community is:

\begin{definition}[Link Community]\label{def:linkcommunity}
A link community in a complex network is a set of nodes that share a number of relations clustered together since they belong to a particular relational environment.
\end{definition}

This approach implies an overlapping partition, since a node belongs to all the communities of its links, and only in rare occasions do all the links belong to a single community. We prove this point in Section \ref{sec:experiments}, by looking at the average number of communities a node belongs to, according to algorithms in this category. One feature that is ignored by this community definition is the direction of a relation, since an undirected link belongs to a single community. There is no way to attach a relationship from $u$ to $v$ to a community and a relationship from $v$ to $u$ to another community, since they both belong to the same relational environment.

The basic approach to the link clustering problem is to define a projection graph in which the nodes represent the links of the original graph and the definition of a proximity value in order to understand how close two edges of the network are. In both cases the critical point is to measure the relations between the edges. A classical clustering algorithm can then be applied.

The methods reviewed here reflect both approaches. The first \cite{link-modularity} defines the projection graph with a random walk measure for the proximity of the projected edges, then uses modularity to compute the modules of the network. The second one \cite{link-jaccard} is a general framework in which it is possible to define any distance measure for the nodes (such as the Jaccard index) and then apply a classical hierarchical clustering technique based on this distance definition. Finally we present also a bayesian approach to this problem \cite{link-bayes}.

\subsection{Link modularity \cite{link-modularity}}
In this work, by defining communities as a partition of the links rather than the set of nodes, the authors interpret the usual modularity $Q$ in terms of a random walker moving on the nodes. They further define two walking strategies: a link-link and a link-node-link random walk. They project the adjacency matrix onto a bipartite incidence matrix. The elements $B_{i\alpha}$ of this $n \times m$ matrix are equal to 1 if link $\alpha$ is related to node $i$, and 0 otherwise.

The incidence matrix is then projected onto a line graph: a link is added between two nodes in this projected graph if these two nodes have at least one node of the other type in common in the original incidence bipartite graph. Modularity is then computed on this line graph. The total complexity of creating the line graph and computing modularity is $\mathcal{O}(2mk\log n)$.

\subsection{Hierarchical Link Clustering HLC* \cite{link-jaccard}}
In this approach, the authors start from the assumption that whereas nodes belong to multiple groups (individuals have families, co-workers and friends), links often exist for one dominant reason (two people are in the same family, work together or have common interests) and therefore they cluster them. They define a link similarity measure as the Jaccard coefficient. This measure is computed on the sets of neighbors of each edge sharing one node (i.e. only adjacent edges). The formula used is:
$$
S(e_{ik},e_{jk}) = \dfrac{|n_{+}(i) \cap n_{+}(j)|}{|n_{+}(i) \cup n_{+}(j)|}
$$
where $e_{ik}$ is an edge between nodes $i$ and $k$ and $n_{+}(i)$ is the set of neighbors of node $i$. The approach can be used with an arbitrary similarity function for the edges. Furthermore, although weights and multipartite structures are not considered with this formula, the authors claim that it is possible to extend the approach in order to obtain such features.

The authors then build a dendrogram with a classical hierarchical clustering approach using the defined similarity measure, with a time complexity of $\mathcal{O}(n\bar{K}^2)$. In the dendrogram each leaf is a link from the original network and branches represent link communities. In the hierarchical structure identified, links occupy unique communities whereas nodes naturally occupy multiple communities, owing to their links. Thus the extracted network structure is both hierarchical and overlapping. The dendrogram is then cut by optimizing the partition density objective function \cite{partition-density}.

\section{Link Maximum Likelihood \cite{link-bayes}}
In this work the general idea of a link clustering is combined with multidimensional networks: the idea is that communities arise when there are different types of edges, i.e. dimensions, in a network. Basically the approach is to generate a model for the observed network with a given partition of edges into link communities and then testing these communities with a maximum likelihood approach. The generation and test is very similar to the technique implemented in the Expectation Maximization \cite{leicht-bayes-def} presented in the following section, but in this case is applied on edges instead of applying it on nodes.

\section{No Definition}\label{sec:lastcategory}
There are a number of frameworks for community discovery that use a very trivial definition of community or have no definition at all. These methods often assume that there are some desirable features of the community that are not provided by many algorithms. They define preprocessing and/or postprocessing operations and then apply them to a number of other different known methods which do not extract communities with the desired features. In this way they improve the results.

Basically, the meta definition adopted is:

\begin{definition}[Community]\label{def:nodefcommunity}
Communities in a complex network are sets which present a number of particular features regardless of why their nodes are grouped together. 
\end{definition}

Of course, the meta procedures and features of these approaches depend on both the pre/postprocess and the ``hosted'' method. The works which present a proper definition of a community are, for instance, the evolutionary clustering \cite{evolutionary-clustering} or the CONGA algorithm \cite{conga2}, which have already been outlined in this survey. Given that we have presented their desired common features for the sets in the form of an independent community definition, we have not included these methods in this category.

Instead we focus on four methods: the first is a hybrid framework combining Bayesian and non-Bayesian approaches \cite{bayeshybrid}, the second relies on a custom definition of community given by the analyst and then performs a multidimensional community discovery, by identifying the noisy relations inside the network \cite{multirelationalhan}, the third one is a bayesian hierarchical approach \cite{clauset-bay-hier}, finally the last one is based on an expectation maximization principle \cite{leicht-bayes-def}.

\subsection{Hybrid* \cite{bayeshybrid}}
For this framework, the authors start from the point that overlapping communities are a more precise description for the multiplicity of node links compared to non-overlapping approaches. If a node's links cannot be explained by a single membership, then the community discovery problem has to be solved in an overlapping formulation. On the other hand, if a node's links can be explained almost equally well by a number of single and mixed memberships, hard clustering may be simpler. The conclusion is that a combination of an overlapping community discoverer that takes an already hard defined community as input with a non overlapping method should perform better. Thus the HFCD framework is built. It is made up of three parts: the Bayesian core, the hint source procedure and the coalescing strategies.

The Bayesian core is the overlapping community discovery algorithm that collects the hints from the other non overlapping method and outputs the final community partition. In \cite{bayeshybrid} the authors use a Latent Dirichlet Allocation on Graphs \cite{lda-bayes-cd,lda-original} as their core method. The Bayesian core needs some hints in order to perform the community discovery procedure. These hints are provided by any other non overlapping community detection algorithm, namely modularity \cite{clauset-modularity} and Cross Associations \cite{cct-base} (here reviewed in its evolution as a Context-specific Cluster Tree \cite{cct}).

The most important contribution of this approach is in creating a procedure that solves the problem of how to incorporate the hints into the core model. This is done by the coalescing strategies. The authors propose three different strategies: attributes (each community is an attribute of the node), seeds (the community partition is used as an initial configuration of the second community discovery phase), and prior (a mix of the previous two). In order to make the inference procedure both for attributes and for the initial configuration, the authors use the Gibbs sampling technique \cite{gibbs-survey}. The additional complexity over the used methods is $\mathcal{O}(nk\bar{K})$.

\subsection{Multi-relational Regression \cite{multirelationalhan}}

\begin{figure}
\centering
\includegraphics[scale=0.4]{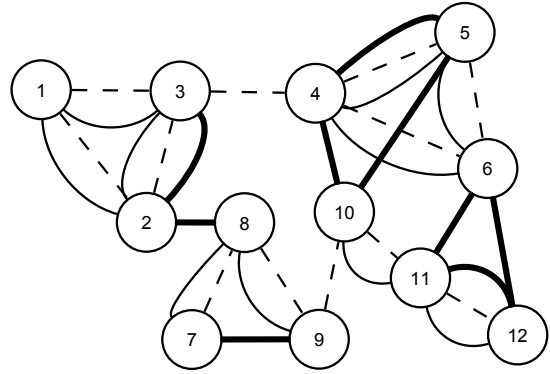}
\caption{A multidimensional network. Solid, dashed and tick lines represent edges in three different dimensions.} \label{fig:multidim-net}
\end{figure}

This algorithm aims to discover hidden multidimensional communities. The authors use the term ``relation'' for a dimension, i.e. a criterion to connect entities. They define relation networks, group them together and create a kind of social network, calling it a multi-relational social network or heterogeneous social network, another name for a multidimensional or multiplex network. The basic assumption is that each relation (explicit or implicit) plays a different role in different tasks.

For instance consider the multidimensional network in Figure \ref{fig:multidim-net}. The authors suppose that an analyst might want to specify that nodes 8, 9, 10 and 11 belong to the same community. The three dimensions (represented by solid, dashed and thick edges) then have a different importance in reflecting the user information needed. The thick dimension can be considered as noise, and the most important dimension is obviously the dashed dimension. The community discovery process should take this situation into account in order to provide an output close to the information needs of the user.

The authors thus represent each relation with a weighted matrix. Each element in the matrix reflects the relation strength between the two corresponding entities. This matrix is then mined depending on a user example (or information need): the user submits a query defining the desired community structure. From this structure, the algorithm reconstructs the possible hidden relation, combining the single relation graphs with linear techniques, and then performs the community discovery on the resulting hidden graph.

The hidden relation is tackled as a prediction problem: once the combination coefficients of the desired entities and the desired relations are computed, the hidden relation strength between any object pair can be predicted. This is a regression problem that can be solved with a number of techniques \cite{regression-resolution}. For a discussion of the issues in this solution based on unconstrained linear regression see \cite{regression-problem}. The exact regression used is the Ridge Regression.

\subsection{Hierarchical Bayes \cite{clauset-bay-hier}}
In this work authors start from the assumption that many real world networks present an hidden hierarchical organization able to explain some of the basic properties of the structure. By reconstructing this latent organization, they are able to group together nodes which are part of the same functional module of the network. It is evident that there is no traditional definition of community at all, and also the authors acknowledge that to reconstruct the hidden dendrogram is a task which goes beyond the simple clustering.

Basically, authors generate and sample a set of dendrograms, which are able to generate a random network with similar features to the observed network, with a Monte Carlo algorithm. The sampling is driven by the maximum likelihood, i.e. the dendrograms are extracted according to how well they can reproduce the observed features. By varying the $p_r$ parameter, the probability to join two vertices in the dendrogram, authors can tune the dendrogram generation in order to fit different properties of the network. Finally, the set of dendrograms is merged into a single consensus dendrogram, which is the best overall representation of the observed network. Although their technique presents an exponential time complexity at the worst case, authors found that in average their complexity should not exceed $\mathcal{O}(n^2)$.

\subsection{Expectation Maximization \cite{leicht-bayes-def}}
This work acknowledge the standard assumption in the community discovery literature, i.e. to define what a community is and then to implement an algorithmic procedure able to create a partition of the network which reflect the best community division according to the starting definition. However, the problem is that sometimes it is hard to define a priori what a community is in a particular network, and failing to do so may end up in finding not significant results. The proposed method is instead able to adapt its definition of community to the most likely present in the data, which may be anyone of the presented classification in this paper.

Basically the authors consider the group membership of each node as an unknown feature. They then define for each vertex $i$ the
probability that a (directed) link from a particular vertex in group $r$ connects to vertex $i$ as $\eta_{ri}$. Finally, $\pi_r$ is the probability of belonging to group $r$. Both $\eta_{ri}$ and $\pi_r$ are unknown and depend on each other. With an iterative, self-consistent approach that evaluates both simultaneously, two characteristic equations which define the expectation maximization algorithm are derived, and the problem can be then solved.

\section{Experiments}\label{sec:experiments}

In this section we briefly present an experimental evaluation of some of the presented algorithms. The aim is to strengthen the intuition regarding the desired features which each category is either able to present naturally or has difficulties with.

\begin{figure}
\centering
\includegraphics[scale=0.4]{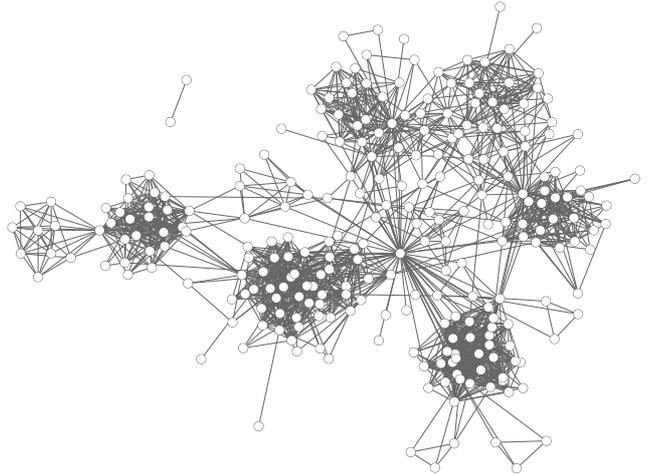}
\caption{Our benchmark network.} \label{fig:exp-network}
\end{figure}

In order to do this, as our benchmark we used a network extracted from the ego network of one of the author's Facebook profiles. We depicted the graph used in Figure \ref{fig:exp-network}. The network contains 261 nodes and 1,722 edges. We chose this network because the human eye can easily spot natural denser areas: there are four main ones at the bottom and left hand side of the picture and three big areas in the upper right hand side, while in the middle there is a sort of gray area and smaller cliques and quasi-cliques of 3-7 nodes float around. We also have a thorough knowledge regarding the nodes and the actual community partition of these people from the perspective of the network ego, however for privacy reasons we cannot include more detailed data.

We have tried to include as many algorithms as possible in this section\footnote{We implicitly thank all the authors of the included algorithms for making them available or sending them to us.}. We excluded reviewed methods for any of the following reasons: we were not able to find any implementation (or working implementation) freely available, the algorithm did not provide better knowledge regarding its category being very similar to another already included, or the algorithm was not suitable for real-world purposes, i.e. it was not able to provide a result on our example network in less than two hours and 1GB of memory occupation (for a 37kB input).

All of the evaluation measures used take a partition P of the network as input, i.e. a list of set of nodes which may or may not have common elements (i.e. overlap).

\begin{table*}
\scriptsize
\begin{center}
\begin{tabular}{|c|rrrrrr|}
\hline
Algorithm & $k$ & $\bar{n}$ & $Q$ & $fl$ & $C^{-1}$ & $o$ \\
\hline
SocDim & 12 & 45.583 & N/A & 6.583 & 0.451 & 2.096\\
Autopart & 6 & 43.500 & 0.309 & 18.500 & 0.212 & 1\\
\hline
Modularity & 8 & 32.625 & 0.724 & 0.375 & 0.744 & 1\\
Local Density & 31 & 8.419 & 0.714 & 0.226 & 0.549 & 1\\
\hline
Edge Betweenness & 11 & 23.727 & 0.738 & 0.455 & 0.656 & 1\\
CONGA & 119 & 5.277 & N/A & 3.958 & 0.076 & 2.406\\
\hline
Label Propagation & 13 & 20.077 & 0.735 & 0.385 & 0.616 & 1\\
\hline
Walktrap & 12 & 21.750 & 0.738 & 0.250 & 0.652 & 1\\
Infomap & 17 & 15.353 & 0.721 & 0.765 & 0.510 & 1\\
\hline
K-Clique & 16 & 16.125 & N/A & 1.562 & 0.341 & 0.989\\
S-Plex & 96 & 3.615 & N/A & 2.417 & 0.070 & 1.330\\
\hline
Link Modularity & 37 & 26.216 & N/A & 3.730 & 0.395 & 3.716\\
HLC & 256 & 3.734 & N/A & 2.539 & 0.063 & 3.663\\
\hline
\end{tabular}
\end{center}
\label{tab:algorithmexp}
\caption{The statistical parameters of communities extracted with different approaches.}
\end{table*}

\begin{itemize}
\item \textbf{Modularity} ($Q$). Although there are overlapping definitions for this measure \cite{nicosia-modularity}, the main version used is the standard one which is not defined for overlapping partitions. Therefore, we computed the original version of Modularity only for non-overlapping results. 
\item \textbf{Flake-ODF} ($fl$), introduced in \cite{leskovec-empirical}, is defined as the fraction of nodes in a community that have fewer edges pointing inside than outside of the cluster. We calculate the average over all communities, i.e. $fl(p) = \sum_{k \in P} \frac{|\{u : u \in k, |\{(u,v):v \in k\}| < deg(u) / 2\}|}{|k|}$. In \cite{leskovec-empirical} many evaluation measures are presented in order to solve the monotonic increase in modularity (i.e. the resolution problem: bigger clusters tend to score better). However, we tested all of them in our experimental setting (some are not reported here for the sake of readability) and we found that all tend to assign constantly lower scores to overlapping partitions in the same network. Thus, these measures should be refined in order to be more general and to include the very common and popular overlap feature. 
\item \textbf{Reverse Conductance} ($C^{-1}$). Conductance is also presented in \cite{leskovec-empirical} as the fraction of total edge volume that points outside the cluster. We are interested in the reverse concept, i.e. the fraction of total edge volume that points inside the cluster, i.e. $C^{-1} = \frac{1}{|P|} \sum_{k \in P} \frac{m_k}{2c_k  + m_k}$, where $m_k = |\{(u,v) \in m : u \in k \wedge v \in k\}|$ and $c_k = |\{(u,v) \in m : u \in k \wedge v \not\in k\}|$. 
\item \textbf{Overlap Ratio} ($o$) is informally defined as the average number of communities that a node belongs to in the network, i.e. $o(p) = \sum_{n \in N}\frac{|\{ k \in P : n \in k \}|}{|N|}$. While a non overlapping community discovery usually returns 1 in this metric, if an algorithm does not cluster all the vertices in the network then it may return a value less than 1.
\end{itemize}
We report the final results in Table 4, in which we have one row per algorithm and one column per measure. We added some statistically simple parameters such as the number of communities and average number of nodes per community. For the measures, in Table 4 we use the same notation used in this section to present them.

We are now able to provide an additional reason for our classification by analyzing the presented results.

SocDim and Autopart belong to the Feature Distance category. As discussed in Section \ref{sec:firstcategory}, in this category we have a method with basically any feature (for example, SocDim is multidimensional and overlapping, while Autopart is parameter free and allows directed edges). The downside is the counter intuitive partition according to the graph topology. It is easy to see, in fact, how poorly Autopart scores in the Modularity test ($Q$). However, since we did not compute Modularity for the overlapping SocDim partition, we also used the Flake-ODF measure ($fl$). In this case too, both SocDim and Autopart got higher values, i.e. it is more frequent that a node has more edges pointing outside the cluster than pointing in. Overlap partitions usually have the lowest performance according to Flake-ODF, and to Conductance, since nodes in the overlap zone are densely connected to two or more clusters. However Autopart is not an overlapping method and SocDim turned out to be the worst of the other overlapping algorithms according to this evaluation.

For the Internal Density category (Section \ref{sec:modularity}) we tested Modularity and Local Density algorithms. Their edge volume inside the community (Reverse Conductance $C^{-1}$) is high. For Modularity edge volume was the highest score, while Local Density scored well, although it did not come second for implementation reasons (the algorithm returns some communities with only one vertex which obviously contributes with zero to the sum).

As stated in Section \ref{sec:bridgecommunity} regarding the bridge detection community discovery, no assumptions about the density of the clusters are made. Thus these algorithms may have a high score on the inverse conductance (Edge Betweenness), or may not (CONGA).

Unfortunately our set of algorithms for the Diffusion (Section \ref{sec:percolation}) category is very narrow and no conclusions can be drawn. Instead, Closeness algorithms (Section \ref{sec:closeness}) Walktrap and Infomap highlight their independence from a simple density definition: Walktrap favors a few bigger (and denser) communities, while Infomap focuses on smaller and lower level sparser ones.

There is one clear downside to the Structure definition category (Section \ref{sec:structure}): the K-Clique algorithm has an overlap ratio $o$ less than one, since its structure definition is very strict and many nodes cannot satisfy it, ending up in no community.

Finally, algorithms in the Link Community category (Section \ref{sec:linkcommunity}) gave a very high overlap score ($o$). This proves that clustering edges is a natural and automatic way to get highly overlapping partitions.

\section{Related Works}\label{sec:related}
Over the last decade, several reviews of community discovery methods have been published. We would consider the most important to be \cite{newmanreview, graphmining, survey3, survey4, portereview, schaeffereview}.

Fortunato and Castellano \cite{survey4}, hugely extended by Fortunato in \cite{bfsurvey}, have published the most recent and probably the most comprehensive review on the community discovery problem. To tackle the problem they consider various definitions of community (local, global and vertex similarity), features of communities for extraction, and different categories. The number of algorithms and references they considered is impressive. We believe that a new review of this topic is needed because the authors analyze the main techniques of each method for community detection, however they do not build an organization of community definitions (while acknowledging that different ones exist). Thus, they do not consider the main contribution of our review: the creation of a classification of community based on definitions of state of the art algorithms. Without focusing on a classification of community definitions, Fortunato and Castellano's review cannot be used by a researcher with his/her own definition of what a community is in order to find the most relevant set of methods for his/her problem. Their review is aimed at people interested in building a new community detection algorithm, not people who want to use the methods in the literature. Furthermore, their work does not include some more advanced features and definitions of community found in the literature, such as multidimensionality or an influence spread formulation of the problem.

Porter et al. \cite{portereview} and Schaeffer \cite{schaeffereview} have also recently reviewed community discovery methods. In \cite{schaeffereview} they also introduced the problem of a comprehensive meta definition of community in a graph. Again, however, although they begin to provide different definitions of community, they do not create a classification of the community discovery algorithm based on such a community.

In Newman's pioneering work \cite{newmanreview} he organizes historical approaches to community discovery in complex networks following their traditional fields of application. He presents the most important classical approaches in computer science and sociology, enumerating algorithms such as spectral bisection \cite{classical-spectral} or hierarchical clustering \cite{sna-book}. He then reviews new physical approaches to the community discovery problem, including the known edge betweenness \cite{edgebetween} and modularity \cite{newman-modularity}. His paper is very useful for a historical perspective, however it records few works and obviously does not taken into account all the algorithms and categories of methods that have been developed since it was published.

Chakrabarti and Faloutsos \cite{graphmining} give a complete survey of many aspects of graph mining. One important chapter discusses community detection concepts, techniques and tools. The authors introduce the basic concepts of the classical notion of community structure based on edge density, along with other key concepts such as transitivity, edge betweenness and resilience. However, this survey is not explicitly devoted to the community discovery problem. It describes existing methods but does not investigate the possibility of different definitions of community or of a more complex analysis.

Danon et al. \cite{survey3} test an impressive number of different community discovery algorithms. They compare the  time complexity and performance of the methods considered. Furthermore, they define a heuristic to evaluate the results of each algorithm and also compare their performance. However, they focus more on a practical comparison of the methods, rather than a true classification, both in terms of a community definition and in the feature considered for the input network.

Various authors have also proposed a benchmark graph, which would be useful to test community discovery algorithms \cite{benchmark}.

\section{Conclusions}\label{sec:conclusions}
The aim of this survey was to create a manual for the community discovery problem, to answer the question: ``Given what is considered a community for analysts, which community detection algorithm should they use?''. This is a sort of orthogonal point of view compared to the classical approach of community discovery reviews, traditionally aimed at analysts already within the community discovery field.

We first tackled the problem of the lack of a universally accepted definition of what is a community. As pointed out by Fortunato \cite{bfsurvey}, this lack of a theoretical framework has some important consequences not only in the community detection task itself (if we do not agree on the meaning of ``community'' how can we extract a community from the network?) but also in other aspects. One of these aspects is, for instance, the evaluation of an algorithm w.r.t. the results from another approach using a different definition of community.

We have proposed a meta definition of community, and on this basis we built a new classification of community discovery methods based on the relationships of each definition of community using the general meta definition. We have reviewed the approaches according to general categories such as internal density, community structure definition and so on. This classification is a proposed answer to the problems highlighted by Fortunato. Each main method is then briefly presented, along with its relationship with other algorithms, its complexity and the strong and weak points of the category it belongs to.

A crucial problem that we have identified is the need for an extensive study of the overlap between the definitions of community. As pointed out in Section \ref{sec:classoverlap} there are several complex connections between different definitions and different algorithms. It would be worth creating an accurate graph representation of this overlap, in which the nodes are the connected algorithms if they share part of their community definition, some features of the input/output, some quality functions or a search space exploration approach. This multidimensional complex network could be studied in order to have a clearer and more detailed view on the community discovery problem.

Another contribution of this paper is the inclusion of the important innovative features of a graph partition algorithm, which has not considered in other reviews. The definition of different features is critical because clearly there is no ``perfect method''. However methods that are or are not able to consider multidimensionality, algorithms that do or do not treat overlapping communities, and so on, can be categorized as such. We have discussed this point in each category, trying to highlight which features are naturally provided by each category and which ones are not. We chose to include novel features like multidimensionality, so far not considered by community discovery reviews, since they add a fundamental analytical power that better describes real world phenomena. Moreover, an approach is not necessarily better if it has a longer list of supported features: in some cases a specialized method can achieve a better performance than a general one. Thus we believe that Table 2 is useful for checking the features of all algorithms. We hope this will help analysts to find the desired algorithm also in terms of features and not only the underlying definition of community.

To define and predict what will be the most important features in the future is another open question that we leave for future work. There is interest especially in multidimensionality \cite{tangcikm, multicd-asonam, onnela, link-bayes, multirelationalhan}, perceived as a feature that is part of the solution and not only as an input to be preprocessed. In other words, we want not only to consider multidimensionality as an input, but also to extract truly multidimensional communities. Another interesting feature might be the presence of both a hierarchical and overlapping organization of the community structure at the same time, since these two features are no longer seen as being mutually exclusive \cite{link-jaccard}.

\textbf{Acknowledgements}. We gratefully acknowledge Sune Lehmann for useful discussions. Michele Coscia is a recipient of the Google Europe Fellowship in Social Computing, and this research is supported in part by this Google Fellowship.

\section{Bibliography}
\bibliographystyle{ieeetr}

\bibliography{stateoftheart}

\end{document}